\documentclass[reprint,aps,pre,amsmath,amssymb,a4paper,superscriptaddress]{revtex4-1}

\usepackage{hyperref}
\hypersetup{
  pdfauthor={Milanese, Brink, Aghababaei, Molinari},
  pdftitle=The role of interfacial adhesion on minimum wear particle size and roughness evolution,
  pdfborder=0 0 0,
  colorlinks=true,
  urlcolor=blue,
  linkcolor=blue,
  citecolor=blue
}
\usepackage{graphicx}
\usepackage{upgreek}
\usepackage{array}
\usepackage{makecell}
\usepackage{float}
\newcolumntype{P}[1]{>{\centering\arraybackslash}p{#1}}

\begin{document}
\sloppy

\title{The role of interfacial adhesion on minimum wear particle size and roughness evolution}

\author{Enrico Milanese}
\affiliation{Civil Engineering Institute, Materials Science and Engineering Institute, \'{E}cole Polytechnique F\'{e}d\'{e}rale de Lausanne (EPFL), CH-1015 Lausanne, Switzerland}
\author{Tobias Brink}
\affiliation{Civil Engineering Institute, Materials Science and Engineering Institute, \'{E}cole Polytechnique F\'{e}d\'{e}rale de Lausanne (EPFL), CH-1015 Lausanne, Switzerland}
\author{Ramin Aghababaei}
\affiliation{Department of Engineering - Mechanical Engineering, Aarhus University, 8000 Aarhus C, Denmark}
\author{Jean-Fran\c{c}ois Molinari}
\email{jean-francois.molinari@epfl.ch}
\affiliation{Civil Engineering Institute, Materials Science and Engineering Institute, \'{E}cole Polytechnique F\'{e}d\'{e}rale de Lausanne (EPFL), CH-1015 Lausanne, Switzerland}

\begin{abstract}

Adhesion between two bodies is a key parameter in wear processes. At the macroscale, strong adhesive bonds are known to lead to high wear rates, as observed in clean metal-on-metal contact. Reducing the strength of the interfacial adhesion is then desirable, and techniques such as lubrication and surface passivation are employed to this end. Still, little is known about the influence of adhesion on the microscopic processes of wear. In particular, the effects of interfacial adhesion on the wear particle size and on the surface roughness evolution are not clear, and are therefore addressed here by means of molecular dynamics simulations. We show that, at short timescales, the surface morphology and not the interfacial adhesion strength dictates the minimum size of wear particles. However, at longer timescales, adhesion alters the particle motion and thus the wear rate and the surface morphology.


\end{abstract}

\maketitle

\section{Introduction}

Wear, the removal of material from interacting surfaces, not only influences the durability of mechanical systems, but is also a source of health concerns. Frictional processes that normally take place when a vehicle is in motion (e.g.\ a car's brakes being pulled together, tyre on pavement) release wear particles in the air. Such airborne particles are known to be a health hazard, as they are associated with an overall increase of death risk~\cite{samet2000fine,pope2002lung}. In particular, the size of the airborne particles has a fundamental role in this, as particles at the nanoscale can deposit in lungs and other organs~\cite{samet2000fine,olofsson2011study}, and several countries prescribe limits to the concentration of fine particles in the air. We thus here explore the role of adhesion on the size of wear particles, and the subsequent effects on the surface morphology, in a simplified two-dimensional setup. Adhesion is in fact known to affect significantly wear debris formation -- most often its effects prevail over those due to other phenomena (e.g.\ corrosion, fatigue)~\cite{rabinowicz1995friction}. In this case, strong bonds develop at the interface between the two surfaces, and, during sliding, bond breaking below one of the two surfaces is favoured. Material is then removed from the solid and it is either transferred to the other surface, or it comes off as a loose wear particle. Loose particles form then the third-body, which alters the system configuration and dynamics~\cite{godet1984third}, before being eventually evacuated from the contact and released into the atmosphere.

Early pioneering work already put forward the concept of a critical contact size for adhesive wear particles to form upon contact~\cite{rabinowicz1995friction}. More recent advances in the understanding of wear lead us to a more complete picture, and we now know that different mechanisms of material transfer are observed within the adhesive wear regime. For low adhesion and light loads, wear follows an Eyring-like atom-by-atom removal mechanism~\cite{bhaskaran2010ultralow,schirmeisen2013wear,jacobs2013nanoscale,yang2016adhesion,shao2017multibond,liu2017tribochemical}. More relevant for particle formation are higher loads and adhesion. In this regime, we now understand that the particle formation criterion~\cite{aghababaei2016critical} is defined by the competition between plastic deformation~\cite{holm1946,merkle2008liquid,yang2016adhesion,aghababaei2016critical} and brittle fracture~\cite{aghababaei2016critical,milanese2019emergence,archard1953contact,liu2010method,liu2010preventing} of the contacting asperities. This transition from ductile to brittle behavior is governed by a material-dependent critical length scale $d^*$~\cite{aghababaei2016critical}. If the junction $d$ formed upon contact by the colliding asperities is smaller than $d^*$, then the asperities deform plastically (Supplementary Figure~\ref{fig:setup}(a)). Vice versa, if $d \ge d^*$, the asperities break, form a debris particle, and the system transitions to a three-body configuration (Supplementary Figure~\ref{fig:setup}(a)). The critical length scale $d^*$ has the form

\begin{align}
  d^* = \Lambda {} \frac{w}{\tau_\mathrm{j}^2 / 2G} \textrm{,}
  \label{eq:dstar}
\end{align}
where $\tau_\mathrm{j}$ is the junction shear strength (affected by the adhesion strength and bulk properties), $G$ is the shear modulus of the material, $w$ is the fracture energy and $\Lambda$ is a geometrical factor (which is of order unity and takes into account the shape of the colliding asperities). This critical length scale $d^*$ explains the resulting transition to a three-body system by a brittle mechanism, which is needed to evolve the initial surface topography into a self-affine morphology~\cite{milanese2019emergence}, and provides further insights into the process of wear debris formation~\cite{aghababaei2017debris,aghababaei2018asperity,Frerot2018,pham2019adhesive,brink2019adhesive}. Furthermore, consistent with the definition of Eq.~\ref{eq:dstar}, it has recently been shown that lower values of interfacial adhesion (i.e.\ lower $\tau_\mathrm{j}$) leads to larger debris volumes upon formation (larger $d^*$), if the initial surfaces are both atomistically flat except for a well-defined asperity~\cite{aghababaei2019effect}.

Yet it is not clear how reductions in the interfacial adhesion strength affect the debris particle formation process for different initial surface morphologies. Frictional surfaces indeed often appear self-affine~\cite{renard2013constant,sayles1978surface,majumdar1990fractal,persson2004nature}, that is they are rough over many length scales. Investigating self-affine surfaces is thus the next natural step following the understanding of the simplified case of well-defined asperities~\cite{aghababaei2019effect}. Moreover, at longer timescales the reduced interfacial adhesion also influences the motion of the debris particles, possibly altering the mechanisms that govern the roughness evolution observed in the full adhesion case~\cite{milanese2019emergence}. Therefore, the present paper is concerned with investigations of self-affine surfaces and of the interplay between their geometry, the interfacial adhesion, and the particle formation and evolution.

\section{Materials and Methods}

The study consists of two sets of molecular dynamics simulations and each set is characterized by a different simulated timescale. In both sets, dry sliding of two opposing two-dimensional (2D) surfaces is investigated, at constant temperature, normal pressure, and sliding velocity. The initial surfaces are self-affine (with Hurst exponent $H$ between $0.3$ and $1.0$) and both consist of the same bulk material. Three different values of interfacial adhesion $\tilde{\gamma}$ are investigated: $\tilde{\gamma} \in \{1.0,0.8,0.6\}$. The interfacial adhesion is expressed in dimensionless terms as $\tilde{\gamma} = \gamma_\mathrm{int} / \gamma_\mathrm{bulk}$, where $\gamma_\mathrm{int}$ is the surface energy of passivated atoms on the surfaces and $\gamma_\mathrm{bulk}$ is the surface energy without any passivation. For $\tilde{\gamma}=1.0$, the full adhesion case is recovered. For $\tilde{\gamma}<1.0$ we speak of reduced interfacial adhesion. During the simulations, atoms belonging to a free surface are detected on the fly and the interaction potential between such atoms is re-assigned to the interfacial adhesion potential characterized by $\tilde{\gamma}$ in order to model passivation of the surfaces~\cite{aghababaei2019effect}. Atom interactions, both in the bulk and between the two different surfaces, are described by the same class of model pair potentials~\cite{aghababaei2016critical,milanese2019emergence}. These potentials allow to explicitly capture at acceptable computational costs the ductile-to-brittle transition in adhesive wear that takes place in the moderate to large adhesion limit~\cite{aghababaei2016critical}. To include such transition within our simulation box, the chosen potential is characterized by $d^*$ smaller than the horizontal box size $l_x$. The short-timescale set of simulations provides insights into the effect on adhesion and random surface topography upon debris particle formation, while the long-timescale simulations allow us to study the effects of adhesion on the long-term surface roughness evolution and on the wear rate.

Throughout the article, quantities are measured in reduced units, the fundamental quantities being the equilibrium bond length $r_0$, the bond energy $\varepsilon$ at zero temperature, and the atom mass $m$.

\vspace{\baselineskip}
\subsection{Interaction potentials}
All simulations belonging to this study are run with scaled versions of the same potential, which belongs to the same class of model pair potentials introduced in Ref.~\onlinecite{aghababaei2016critical} and also used in Ref.~\onlinecite{milanese2019emergence}.
This family of model potentials is a modified version of the Morse potential~\cite{morse1929diatomic}:

\begin{align}
  \frac{V(r)}{\varepsilon} = \zeta {} \begin{cases} (1-\mathrm{e}^{-\alpha(r-r_0)})^2-1 & r<1.1r_0 \\
    c_1\frac{r^3}{6} + c_2\frac{r^2}{2} + c_3r + c_4 & 1.1r_0 \le r \le r_\mathrm{cut} \\
    0 & r_\mathrm{cut} \le r \end{cases} \textrm{ ,}
\end{align}
where $\zeta$ is a scaling factor that equals 1 for bulk atoms and can be smaller than 1 for surface atoms, $r$ is the distance between two atoms, $\varepsilon$ is the bond energy at zero temperature, $r_0$ is the equilibrium bond length, and $\alpha=3.93$~$r_0^{-1}$ governs the bond stiffness. The $c_i$ coefficients are chosen such that the potential $V(r) / \varepsilon$ is continuous both in energy and force. The cut-off distance is set by $r_\mathrm{cut}$ and determines the inelastic behavior. This allows for changes in the potential tail (and, thus, in the material yield strength), while keeping the same elastic properties up to a $10\%$ bond stretch. The potential adopted in this study to model the bulk is characterized by $r_\mathrm{cut}=1.48$~$r_0$. In Refs.~\onlinecite{aghababaei2016critical} and~\onlinecite{milanese2019emergence} this potential is called P4. The interfacial potentials, i.e.\ the potentials used to represent the adhesion between passivated atoms, are scaled versions of the bulk potential. Three scaling factors $\zeta$ are used, which corresponds to the three different $\tilde{\gamma}$ investigated: $1.0$ (full adhesion), $0.8$, and $0.6$. The on-the-fly algorithm to assign surface atoms to the interfacial adhesion potential is the same adopted in Ref.~\onlinecite{aghababaei2019effect} and works as follows. At time $t=0$ all the atoms belong to the bulk potential. All atoms in the simulation box (except those where a thermostat or displacements are prescribed) are then checked every $1\,000$ time steps: if their coordination number is less or equal to $n_c$, they are considered to belong to a free surface and thus passivated and are re-assigned to the interfacial potential chosen for that simulation. We used $n_c=15$ (within a radius of $2.23$~$r_0$) as in Ref.~\onlinecite{aghababaei2019effect}.

\vspace{\baselineskip}
\subsection{Simulation geometry and boundary conditions}

All simulations were performed in 2D using the molecular dynamics simulator LAMMPS~\cite{plimpton1995fast}. A simple scheme of the simulation setup is shown in Supplementary Figure~\ref{fig:setup}. Two different horizontal box sizes have been adopted, i.e.\ $l_x = 339.314$~$r_0$ (sets S and L) and $l_x = 678.627$~$r_0$ (set L, see below). Periodic boundary conditions are enforced along the horizontal direction. The initial vertical box size is the same for all simulations of all sets and is $l_y = 394.823$~$r_0$, the box is then allowed to expand vertically, e.g.\ upon debris particle formation. A constant pressure ($f_y = 0.02$~$\varepsilon r_0^{-2}$) is applied on the top and bottom boundaries to press the surfaces together and avoid that the surfaces are driven away by inertia at the first collision. A constant horizontal velocity $v_\mathrm{ref}=0.01$~$\sqrt{\varepsilon m^{-1}}$ is imposed on the first layer of atoms of the top surface. The bottom layer of atoms of the bottom surface is fixed. A temperature of $0.075$~$\varepsilon$ (expressed in terms of equivalent kinetic energy per atom) is enforced by means of Langevin thermostats with a damping parameter of $0.05$~$r_0/\sqrt{\varepsilon m^{-1}}$. On each body, the thermostats are applied to the three layers of atoms next to the layer where the fixed displacement or velocity is imposed. The time integration is performed with a time step of $0.005$~$r_0/\sqrt{\varepsilon m^{-1}}$. The two sets differ in the duration of the simulated timescale: set L contains long timescale simulations, for a total of minimum $1.200$~billion and maximum $3.113$~billion time steps, while all simulations in set S are run for $40$ million time steps. The main features of the two sets are summarized in Supplementary Tables~\ref{tab:S10},~\ref{tab:S08},~\ref{tab:S06}, and~\ref{tab:L}. The starting geometry of the system is obtained by filling the whole simulation box with atoms relaxed at the target temperature, and then removing a subset of them based on a purely geometric criterion to obtain two distinct rough surfaces. The self-affine morphology is generated with a random phase filter~\cite{jacobs2017quantitative}.

\vspace{\baselineskip}
\subsection{Self-affine surfaces}

The same definitions and conventions of Ref.~\onlinecite{milanese2019emergence} are adopted throughout the manuscript and are briefly summarized here. For more details on fractal concepts, see Refs.~\onlinecite{barabasi1995fractal} and~\onlinecite{meakin1998fractals}.

Fractal surfaces whose heights $h(x)$ scale differently than the horizontal distance $x$ are self-affine fractals, and they obey the scaling relation $h \left( \xi x \right) \sim \xi^{H} h \left( x \right)$~\cite{barabasi1995fractal}, where $\xi$ is the scaling factor and $H$ is the Hurst (or roughness) exponent, with $0<H<1$ for fractional Brownian motion (fBm)~\cite{mandelbrot1985self}. The Hurst exponent describes the correlation between two consecutive increments in the surface: if $H=0.5$, the increments are randomly correlated (i.e.\ standard Brownian motion), if $0<H<0.5$, the increments are negatively correlated, and if $0.5<H<1$ the increments are positively correlated.

The generation of engineering surfaces is non-stationary and random~\cite{sayles1978surface}, and it can be described as a non-stationary process with stationary increments. This allows to relate the fractal dimension $D$ of the surface with its Hurst exponent $H$ through its Euclidean dimension $n$~\cite{mandelbrot1985self}: $ D + H = n + 1$. For this class of surfaces, assuming a 1D surface profile, the fractal dimension $D$ and the power law exponent $\alpha$ of the power spectral density are related as $\alpha = 5 - 2D$~\cite{ganti1995generalized,berry1980weierstrass}. Under these assumptions, a direct relation between $H$ and $\alpha$ is found: $H = \left( \alpha - 1 \right)/2$.

\vspace{\baselineskip}
\subsection{Surface analysis}

The power spectral density (PSD) of a 1D surface $h(x)$ in terms of PSD per unit length $\Phi_h(q)$, $q$ being the wavevector, is defined as~\cite{press2007numerical}

\begin{align} \label{eq:psdu}
  \Phi_h\left( q \right) \equiv \frac{1}{L} \left| \int_{L} h \left( x \right) \mathrm{e}^{-\mathrm{i}qx} \textrm{d}x \right|^2  \textrm{ ,}
\end{align}
where the integral is the continuous Fourier transform of $h(x)$ and $L$ is the surface length projected on the horizontal axis $x$. The surface profile $h(x)$ is a continuous function and it contains the value of the surface height at each value of the spatial coordinate $x$.
In particular, we estimate $\Phi_h$ as

\begin{align} \label{eq:rel}
  \Phi_h \left( q_n \right) \approx \Delta x {} P_h \left( q_n \right)\textrm{ ,}
\end{align}
where $P_h \left( q_n \right)$ is the classical periodogram~\cite{vanderplas2018understanding,press2007numerical}:

\begin{align} \label{eq:periodogram}
  P_h \left( q_n \right) = \frac{1}{N} \left| \sum_{k=0}^{N-1} h_k \mathrm{e}^{-\mathrm{i} q_n x_k} \right|^2 \textrm{ ,}
\end{align}
the summation being the discrete Fourier transform of the surface. In fact, $h(x)$ is known only at a discrete set of $N$ points $x_k$ ($k=0,1,\dots,N-1$), regularly sampled at an interval $\Delta x$, such that $h_k=h(k \Delta x)$ are the known values of $h(x)$. In our case, $\Delta x = L / N$, $N$ being the number of atoms belonging to the surface of length $L$ ($\Delta x \approx 1$~$r_0$).

The Hurst exponent $H$ can also be estimated with the height--height correlation function~\cite{barabasi1995fractal}, which describes the average change of heights $\Delta h$ between two points at a horizontal distance $\delta x$:

\begin{align}
  \Delta h (\delta x) = \langle \left[ h(x+\delta x) - h(x) \right]^2 \rangle ^{1/2} \textrm{ ,}
\end{align}
where the angle brackets indicate spatial average. $H$ can be derived by the log--log plot of $\Delta h (\delta x)$, as the height-height correlation function scales as $\Delta h (\delta x) \sim \delta x^H$.

The surface roughness, that is the variations in height of the surface profile with respect to an arbitrary plane of reference~\cite{bhushan2000modern}, is measured here in terms of the equivalent root mean square of heights

\begin{align}
  \sigma_\mathrm{eq} = \sqrt{ \sigma_\mathrm{top}^2 + \sigma_\mathrm{bottom}^2 } \textrm{ ,}
\end{align}
where $\sigma_\mathrm{top}$ and $\sigma_\mathrm{bottom}$ are the root mean square of heights of the top and bottom surface respectively. The root mean square of heights $\sigma$ of a surface profile $h(x)$ is defined as:

\begin{align}
  \sigma = \sqrt{ \frac{1}{N} \sum_{k=1}^N h_k^2 } \textrm{ ,}
\end{align}
where $N$ is the number of discretization points of the surface and $h_k$ is the distance of the point $k$ from the plane of reference (the surface mid-plane, in our case).

\vspace{\baselineskip}
\subsection{Data analysis}

All the simulations were visualized with OVITO~\cite{stukowski2009visualization}. Due to the large amount of data, frames were saved every $10^6$~steps for both the simulations in set L and set S.

In order to obtain the surface morphology and the debris particle volume, we need to define a way to differentiate the particle from the surfaces, since the particle is usually in contact with at least one surface. We start from the assumption that the interface between particle and surface should probably have a minimal length. As such, we chose a simulated annealing approach \cite{Kirkpatrick1983} using a semi-grand-canonical lattice Metropolis-Monte Carlo \cite{Metropolis1953} algorithm to find this minimal interface by penalizing the interface as described below. We found that to reliably detect the particle without manual intervention, some heuristics for the initial state are needed. As a preparation step, we therefore split the system into horizontal bins of height $7.5\,r_0$ and define those with a number density of atoms below 70\% of the bulk value as likely regions for the gap between surfaces (which should contain the rolling particle). The atoms above this gap are assigned type ``1'' and below the gap type ``2''. The very bottom and top bins are fixed to never change their type. Now two simulated annealing runs are started, each with pseudo-temperatures of $k_BT = 2.1$, $0.9$, $0.3$, down to $0.0$. The difference between these two runs is that we switched all atoms in the gap to either type 1 or type 2 initially. We found that this works better than to use three types, one of which represents the debris particle. The Monte Carlo trial moves were performed as follows:
\begin{enumerate}
\item Choose a random atom $i$ that has at least one unlike neighbor and switch its type ($1 \rightarrow 2$ or $2 \rightarrow 1$).
\item Calculate the pseudo-energy difference $\Delta E = E_\text{after} - E_\text{before}$, with
  \begin{equation}
    E = \sum_{i=1}^{N_\mathrm{tot}} \sum_{j \in \text{NN}_i}
    \begin{cases}
      0.0 & T_i = T_j \\
      0.8 & T_i \neq T_j \text{ and } j \in P \\
      1.0 & T_i \neq T_j \text{ and } j \notin P
    \end{cases},
  \end{equation}
  where $N_\mathrm{tot}$ is the number of atoms, NN$_i$ is the set of atoms in the first nearest neighbor shell of $i$, $T_i$ is the type of atom $i$, and $P$ is the set of passivated atoms. Note that this pseudo-energy is a purely numerical parameter for the optimization algorithm. The reduced bond energy for passivated atoms favors interfaces along previously passivated regions, which we regard as the most sensible demarcations of the debris particle.
\item If $\Delta E \le 0$ accept the trial move, else accept the trial move with probability $\exp(-\Delta E / k_BT)$.
\end{enumerate}
For one Monte Carlo step, $N_\mathrm{tot}$ of these trial moves are performed. For the two high temperatures, 50 Monte Carlo steps are performed and for the two low temperatures 200. For each simulated annealing run, the lowest energy configuration is saved. These two configurations are compared, and the atoms that differ in type between the two are marked as belonging to the debris particle. This works because the two sub-surface bulk regions (top and bottom) keep their respective type 1 or 2 in both runs, while the atom types in the debris particle will depend on the initial state and thus be different in both runs. We verified by visual inspection of all simulations that the algorithm performs adequately.

Once the debris particle is identified, the surfaces are reconstructed by identifying atoms with a coordination number smaller than 15. For each surface of each analyzed time step, a bijective profile is reconstructed by linear interpolation of the $N$ surface atoms. The surface profile is then discretized in $N$ equally spaced points~\cite{milanese2019emergence}. The surface analysis is performed on these discretized reconstructions of the surfaces. The PSD and structure functions are averaged over 15 samples, spaced such that the debris particle rolls over the whole surface at least one time between two consecutive samples. Data for $\sigma$ and $\sigma_\mathrm{eq}$ is averaged over $10$ consecutive data points.

The tangential force values for simulations in set S are stored every $5\,000$~steps and are averaged over windows of $10^6$~steps to have the same discretization of the volume detection algorithm (see Figure~\ref{fig:S_V0}(d)).

The debris particle volume $V$ and its initial volume $V_0$ are computed by multiplying the number of atoms belonging to the debris particle and the atomic volume~\cite{aghababaei2017debris,milanese2019emergence}, which is $\sqrt{3}/2$~$r_0$.

\section{Results}

In the following sections we discuss the results of our investigations. In section~\ref{ssec:particle}, we find that the initial random morphology governs the initial debris particle volume, and deviations in the adhesion strength from the full adhesion regime do not have a significant influence. The minimum wear particle size in the general situation of rough surfaces is then predicted by the critical length scale of Eq.~\ref{eq:dstar}, with the junction shear strength given by the full adhesion case (i.e.\ $\tau_\mathrm{j} = \tau_\mathrm{bulk}$). At long timescales, we find that the surface morphology is self-affine with a persistent Hurst exponent, provided that the adhesion is strong enough to ensure the continuous re-working of the surfaces due to the third body (section~\ref{ssec:long}), and that the reduced interfacial adhesion affects the wear particle growth (section~\ref{ssec:wearrate}).

\vspace{\baselineskip}
\subsection{Particle formation}\label{ssec:particle}
\begin{figure}
  \centering
  \includegraphics[width=0.46\textwidth]{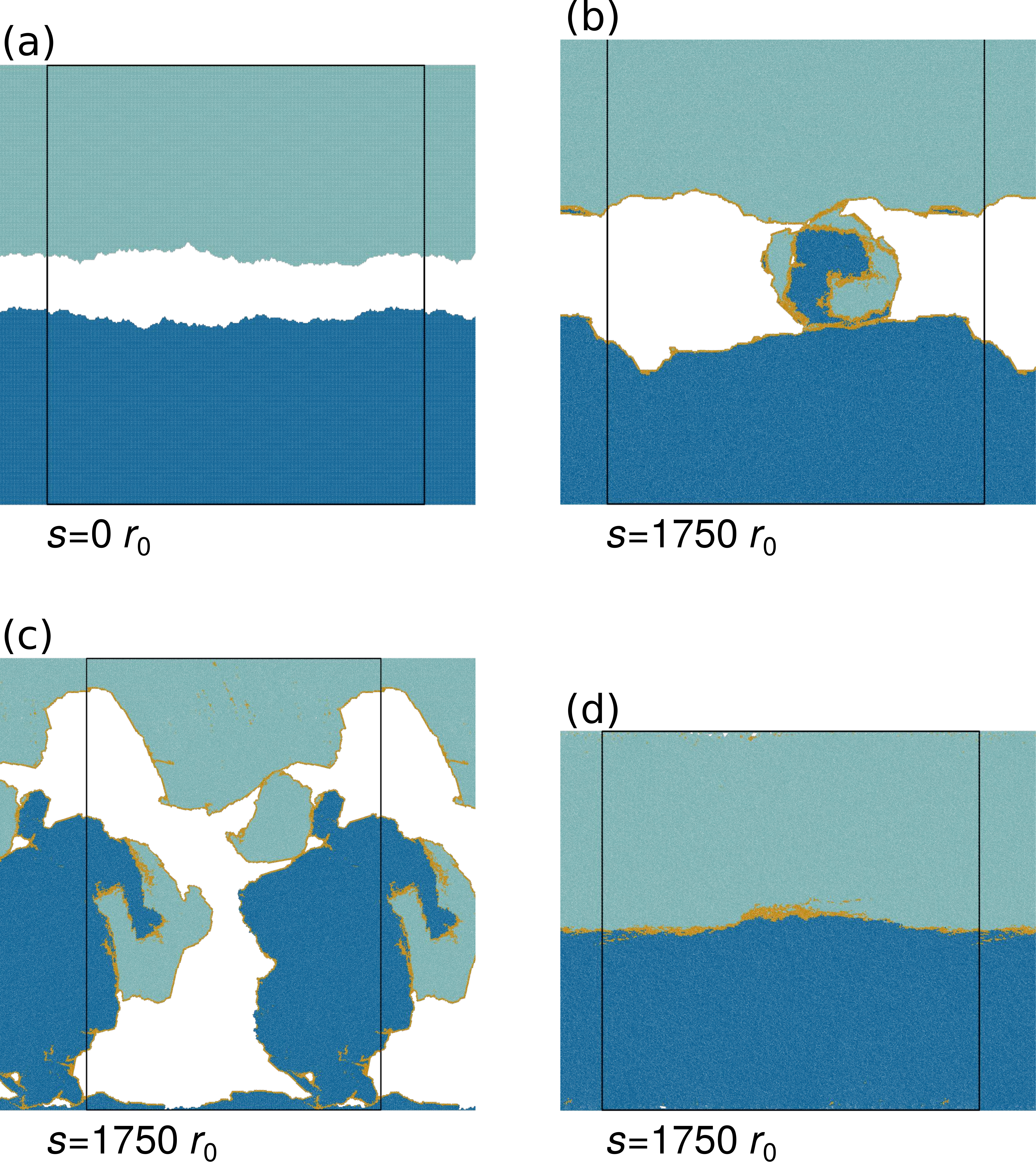}
  \caption{Short timescale evolution. Starting from the same geometry (a), two different scenarios can develop. Either a well-defined debris particle is formed (Scenario~1, (b), frame from simulation S-100-01, see Supplementary Table~\ref{tab:S10}), or the effective contact spreads throughout the interface. In the latter case (Scenario~2), the system attempts to create a debris particle comparable in size with $l_x$ (c), frame from simulation S-080-01, see Supplementary Table~\ref{tab:S08}), or even larger, which results in welding of the interface within the simulated box size and damage initiating near the boundaries (d), frame from simulation S-060-01, see Supplementary Table~\ref{tab:S06}). In all panels colors distinguish atoms originally belonging to the top (light blue) and bottom (dark blue) surfaces. Atoms that at some previous instant were detected as surface atoms and re-assigned to the interfacial adhesion potential are depicted in yellow. Black lines represent simulation box boundaries and $s$ is the sliding distance.}
  \label{fig:S_evo_V0}
\end{figure}

We first focus on the effect of surface morphology and reduced interfacial adhesion in the early stage of the adhesive wear process, when the wear debris particle is formed. We thus ran $135$ short-timescale simulations (set S) with a sliding distance of  $2\,000$~$r_0$. Within this set, we explored different values of the interfacial adhesion ($\tilde{\gamma} \in \{1.0,0.8,0.6\}$), the initial Hurst exponent ($H \in \{ 0.5, 0.7, 1.0\}$) and the root mean square of heights ($\sigma \in \{ 5, 10, 20\}$~$r_0$) of the surfaces. For each value of $(H, \sigma)$, five different random seeds are used to generate five different initial fractal surfaces. The simulations of this set are identified with the first letter 'S' (where 'S' stands for 'short timescale'), followed by three digits that are representative of the value of the interfacial adhesion $\tilde{\gamma}$, and a progressive two-digit number 01 to 45 that is linked to a set of values $(H, \sigma, \textrm{seed})$: simulation S-080-01 thus indicates a short timescale simulation, with $\gamma=0.8$ and $(H, \sigma, \textrm{seed})= (0.5, 5\textrm{~}r_0, 19)$. Details are reported in Supplementary Tables~\ref{tab:S10},~\ref{tab:S08}, and~\ref{tab:S06}. The simulated timescale is large enough to fully reproduce the debris particle formation (see Figure~\ref{fig:S_evo_V0}) and obtain the initial debris particle volume $V_0$.

We observe that at the beginning of the sliding process, when two rough surfaces come into contact, the formation of a wear particle is complex and not well defined: contact can develop at multiple spots along the surface, different contact junctions interact elastically~\cite{aghababaei2018asperity,pham2019adhesive}, and they can coalesce into fewer, larger junctions. The process of debris particle formation is then markedly affected by the surface topography and hard to predict. This situation is more complex than the simplified case of a system with two non-random surfaces, e.g.\ two atomistically flat surfaces exhibiting each a well-defined semicircular asperity. In such case, a contact junction is clearly formed only along the contact interface of the two asperities and Eq.~\ref{eq:dstar} fully describes the loose particle formation. When the two asperities come into contact, they either form a junction of size $d \ge d^*$ and create a wear particle immediately, or they form a junction $d < d^*$, which, upon continuous sliding, increases until $d = d^*$ and a debris particle is formed~\cite{aghababaei2016critical}. (If the asperities are not large enough, $d$ remains smaller than $d^*$ and the two asperities mutually deform plastically, until the surfaces are smooth enough and welding of the interface takes place~\cite{aghababaei2016critical}).

Within the 135 simulations, two different scenarios are observed. Scenario~1: the initial collisions lead to the formation of a distinct wear debris particle -- this is observed in $55.6\%$ of the cases ($75$ simulations). Scenario~2: multiple interacting contact junctions form or the contact spreads throughout the whole system, and a debris particle of characteristic size $d \sim l_x$ or larger would be formed. In such cases, cracks propagate from the surface until they reach the boundaries of the system (Figure~\ref{fig:S_evo_V0}(c)) or the surfaces weld at the interface (Figure~\ref{fig:S_evo_V0}(d)). In the latter case, the periodic boundary conditions suppress any stress concentration required for crack propagation, and a larger system size would be needed to observe cracks that lead to debris particle formation. Because in Scenario~2 cracks either reach the boundaries of the simulation cell (Figure~\ref{fig:S_evo_V0}(c)) or are inhibited (Figure~\ref{fig:S_evo_V0}(d)), simulations that display such scenario are discarded from the analysis. Scenario~2 is observed in the remaining $44.4\%$ ($65$ simulations).

The likelihood of one scenario or the other correlates with the root mean square of heights $\sigma$ of the initial surfaces (see Supplementary Table~\ref{tab:V0}). This is due to the fact that, for a given value of $\tilde{\gamma}$, the rougher the surface, the more pronounced the asperities and valleys, and the system is more likely to form a junction size smaller than the system size $l_x$ (Scenario~1). When surfaces are smooth, the probability of having multiple contact spots that interact and/or coalesce is larger, and the system is more likely to attempt to create a particle of characteristic size $d \sim l_x$ or larger (i.e.\ Scenario~2).

Scenario~1 is also observed more often when adhesion is larger (see Supplementary Table~\ref{tab:V0}). A reduction in $\tilde{\gamma}$ is expected to reduce the junction shear strength $\tau_\mathrm{j}$ and thus to increase the critical length scale $d^*$ (see Eq.~\ref{eq:dstar}). As the system needs a larger junction size to create a debris particle, contact can develop at other places at the same time, increasing the likelihood of Scenario~2.

Note that when the surface roughness and the interfacial adhesion are minimum, i.e.\ $\tilde{\gamma}=0.6$ and $\sigma=5$~$r_0$, no simulation displayed Scenario~1, consistent with the two effects that were just described.

\begin{figure*}
  \centering
  \includegraphics[width=\textwidth]{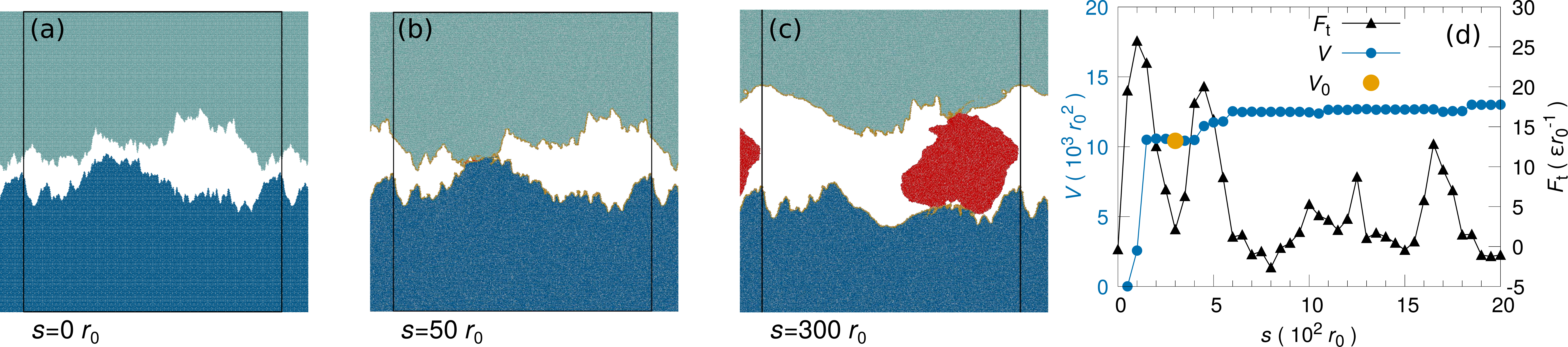}
  \caption{Scenario~1: debris particle formation and initial volume $V_0$. (a-c) The surfaces, initially self-affine (a), come into contact at multiple points (b) and a peak in the tangential force $F_t$ is recorded (d). Upon further sliding, the contact junction is large enough to generate a debris particle, whose formation is over (c) when the first local minimum of the tangential force $F_t$ is reached (d). At this moment the initial debris particle volume $V_0$ is measured (d). (d) Recorded tangential force $F_t$ (black triangles), measured debris particle volume $V$ (blue circles) and measured initial debris particle volume $V_0$ (large orange circle) during a simulation that exhibits Scenario~1. In panels (a-c) colors distinguish atoms originally belonging to the top (light blue) and bottom (dark blue) bodies; in panels (b-c) colors further identify atoms that at some previous instant were detected as surface atoms and re-assigned to the interfacial adhesion potential (yellow) and atoms detected as belonging to the debris particle (red); in panels (a-c) black lines represent simulation box boundaries. In all panels $s$ is the sliding distance expressed in units of $r_0$. Snapshots in panels (a-c) and data in panel (d) are from simulation S-060-32 (see Supplementary Table~\ref{tab:S06}).}
  \label{fig:S_V0}
\end{figure*}

For the subset of simulations that exhibit Scenario~1, i.e.\ a debris particle smaller than the system size is formed, the initial volume $V_0$ is investigated. For this, an unambiguous definition of $V_0$ is needed, and we use the tangential force $F_\mathrm{t}$ as a reference. When the two surfaces first come into contact, the tangential force $F_\mathrm{t}$ starts increasing. After a peak is reached, the force decreases, signaling debris particle formation and the onset of rolling (Figure~\ref{fig:S_V0}). We thus define $V_0$ as the volume of the debris particle measured at the first local minimum exhibited by $F_\mathrm{t}$, after the initial peak, as it corresponds to the work needed to form the debris particle~\cite{aghababaei2017debris}. Identified volumes are then checked for erroneous measures, which are discarded. Erroneous measures are due to false positives of the particle detection algorithm and to the tangential force exhibiting a peak and a following local minimum when the particle is not formed yet (e.g.\ because of an initial ductile event). This procedure allows us to compare consistent values of $V_0$ from the different simulations.

\begin{figure}
  \centering
  \includegraphics[width=0.49\textwidth]{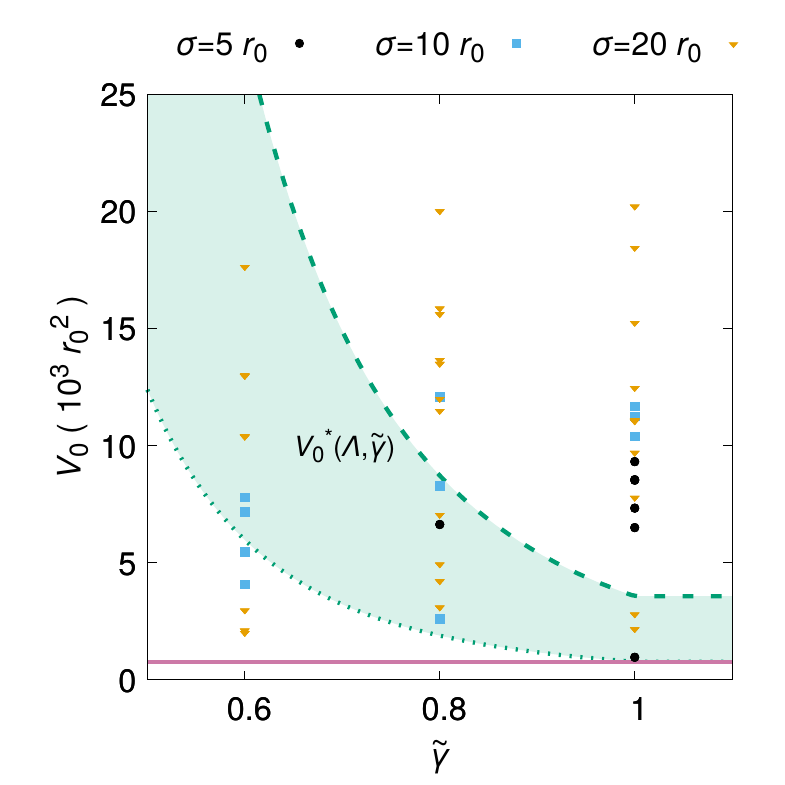}
  \caption{Effect of interfacial adhesion $\tilde{\gamma}$ and surface morphology on the initial debris particle volume $V_0$. In all cases $V_0$ is larger than the minimum size $V_{0,\mathrm{min}}^*$ determined by the material critical length scale (solid purple line). The actual value $V_0 \ge V_{0,\mathrm{min}}^*$ is controlled by the random morphology. No statistically significant increase in the minimum $V_0$ is observed when the interfacial adhesion decreases, as would be expected by a corresponding reduction in the junction shear strength (green shaded area). The green dotted line corresponds to $V_0^*(\Lambda_\mathrm{min},\tilde{\gamma})$ and the green dashed line to $V_0^*(\Lambda_\mathrm{max},\tilde{\gamma})$. Each symbol identifies a unique initial surface roughness in terms of root mean square of heights $\sigma$: $\sigma=5$~$r_0$ (black circles), $\sigma=10$~$r_0$ (light blue squares), and $\sigma=20$~$r_0$ (orange triangles). One data point ($\tilde{\gamma}=1.0$, $V_0=40\,339$~$r_0^2$, $\sigma=5$~$r_0$) is not represented for readability.}
  \label{fig:S_all_V0}
\end{figure}

The data for $V_0$ is reported in Figure~\ref{fig:S_all_V0} and Supplementary Tables~\ref{tab:S10},~\ref{tab:S08},~\ref{tab:S06}, and~\ref{tab:V0}. We observe (Supplementary Table~\ref{tab:V0}) that the average value of the initial volume $\overline{V}_0$ increases with $\tilde{\gamma}$, contrary to Eq.~\ref{eq:dstar}. While at first this seems surprising, as lower values of $\tilde{\gamma}$ imply larger $d^*$, we argue that this is an artifact of the system size for low values of $\tilde{\gamma}$. When reducing $\tilde{\gamma}$, more simulations display in fact Scenario 2, i.e.\ the system attempts to form a debris particle that is too large with respect to the simulation cell. These cases are then not captured by the values of $\overline{V}_0$ that we measured.

When identical initial geometries are compared (i.e.\ for the same values of $(H, \sigma, \mathrm{seed})$), no correlation between the strength of adhesion and the initial volume $V_0$ arises. This leads to the observation that the randomness of the initial surface morphology governs the debris particle formation, in contrast to what is observed at long timescales, where the effect of adhesion is significant (see \textit{Wear rate} paragraph).

We now investigate the effects of the interfacial adhesion and the morphology on the minimum size of the generated debris particle. We know that the critical length scale $d^*$ governs the particle formation process, and that the strength of the interface enters the definition of $d^*$ by affecting $\tau_\mathrm{j}$: Once a junction of size $d<d^*$ is formed, if the interface is weak the asperities slide against one another, otherwise they deform plastically~\cite{brink2019adhesive}. We thus rewrite Eq.~\ref{eq:dstar} to make explicit the effect of the interfacial adhesion,

\begin{align}
  d^*(\Lambda, \tilde{\gamma}) = \Lambda {} \frac{2Gw}{\left( \tilde{\gamma} {} \tau_\mathrm{j,full} \right)^2} \textrm{,}
  \label{eq:dstar_gamma}
\end{align}
where at the denominator the junction shear strength in the case of reduced interfacial adhesion is expressed as a reduction of the junction shear strength $\tau_\mathrm{j,full}$ of the full adhesion case ($\tilde{\gamma}=1.0$), with the proportionality given by $\tilde{\gamma}$ for the potentials adopted in this work~\cite{aghababaei2016critical}. In the limiting case of atomistically flat surfaces with semicircular asperities, this correctly predicts wear particle volumes that are larger when the interface is weaker~\cite{aghababaei2019effect}. To verify such prediction in the case of random rough surfaces, we assume that the minimum initial volume $V_0^*$ is given by a circular particle (in two dimensions), i.e.\

\begin{align}
  V_0^*(\Lambda, \tilde{\gamma}) = \frac{\pi}{4} {} d^*(\Lambda, \tilde{\gamma})^2  \textrm{.}
  \label{eq:Vstar_gamma}
\end{align}
$V_0^*$ is then minimum when $\Lambda$ is minimum and $\tilde{\gamma}$ is maximum, and vice versa. In our work, the extreme values of $\tilde{\gamma}$ are the minimum and maximum input values, i.e.\ $\tilde{\gamma}_\mathrm{min}=0.6$ and $\tilde{\gamma}_\mathrm{max}=1.0$. The geometrical factor $\Lambda$ is instead determined by the geometry of contact, which is hard to obtain in our case of self-affine surfaces. The two limiting cases that we assume for our simulations are flat contact ($\Lambda_\mathrm{min} = 0.70$, see Supplementary Methods)~\cite{pham2019adhesive} and contact between well-defined semicircular asperities ($\Lambda_\mathrm{max} = 1.50$)~\cite{aghababaei2016critical,brink2019adhesive}. If the reduced interfacial adhesion would play a key role in the initial debris particle volume $V_0$, we would thus expect that

\begin{align}
  V_0 \ge V_0^*(\Lambda, \tilde{\gamma}) \textrm{,}
  \label{eq:Vstar_exp}
\end{align}
where $V_0$ is the initial volume observed in our simulations and which is reported in Figure~\ref{fig:S_all_V0}. The green shaded area in Figure~\ref{fig:S_all_V0} depicts the values of $V_0^*(\Lambda, \tilde{\gamma})$, where the limiting cases of $\Lambda = \Lambda_\mathrm{min}$ and $\Lambda = \Lambda_\mathrm{max}$ are given by the lower green dotted line and the upper green dashed line, respectively. Our results show that the initial debris particle volume $V_0$ does not obey the trend of Eq.~\ref{eq:Vstar_exp}, as we observe volumes $V_0$ measured for $\tilde{\gamma}=0.6$ that are smaller than the expected lower bound $V_0^*(\Lambda=\Lambda_\mathrm{min}, \tilde{\gamma}=\tilde{\gamma}_\mathrm{min})$.

This shows that changes in the interfacial adhesion do not affect significantly the initial volume $V_0$ of the debris particle. Instead, we observe that all the recorded values of $V_0$ are larger than $V_{0,\mathrm{min}}^*=V_0^*(\Lambda_\mathrm{min}, \tilde{\gamma}_\mathrm{max})$ (solid purple line in Figure~\ref{fig:S_all_V0}), determined from the minimum $d^*$ given in the full adhesion case ($d^*_\mathrm{full}=d^*(\Lambda_\mathrm{min}, \tilde{\gamma}_\mathrm{max})=31.40$~$r_0$, with the values of $G$, $w$, $\tau_\mathrm{j,full}$ as in Ref.~\cite{aghababaei2016critical}).  We ascribe such behavior to the morphology of the surfaces, which in the current work is random. Recently, a refined version of Eq.~\ref{eq:dstar} was derived~\cite{brink2019adhesive}, where the junction shear strength depends on the angle of contact between the two ideal asperities. It was shown that if the angle of contact is larger than a critical value, the junction shear strength $\tau_\mathrm{j}$ is given by $\tau_\mathrm{j,full}$ and independent of $\tilde{\gamma}$. In the case of contact between self-affine surfaces, the contact junction is rough and the concept of angle of contact is ill-defined. Interlocking is thus expected in at least some cases. Furthermore, we argue that the ratio between the contact size and the thickness of the passivated layer also plays a significant role in the value of $\tau_\mathrm{j}$ and, thus, of $V_0^*$. In our simulations, the junction size is much larger than the passivated layer, which is thin when the surfaces come into contact. The bulk strength thus markedly affects the junction strength $\tau_\mathrm{j}$, which can be approximated by $\tau_\mathrm{j,full}$. If the thickness of the passivated layer were larger than the contact size, then we would expect $\tau_\mathrm{j}$ to be influenced by $\tilde{\gamma}$ and thus $d^*$ to be close to the reduced adhesion value $d^*(\Lambda, \tilde{\gamma}) > d^*_\mathrm{full}$. (Note that in such case a proportional reduction in the fracture energy $w$ is also expected, but the effect on $\tau_\mathrm{j}$ is squared and thus $d^*(\Lambda, \tilde{\gamma}) > d^*_\mathrm{full}$.)

We also recall that for the smoothest cases ($\sigma = 5$~$r_0$), the likelihood of Scenario~1 correlates with the strength of the interfacial adhesion (Supplementary Table~\ref{tab:V0}). This is consistent with the aforementioned argument that when the contact interface is thin enough (compared with the passivated layer), $\tilde{\gamma}$ affects the minimum debris particle volume (Eq.~\ref{eq:dstar_gamma} and~\ref{eq:Vstar_gamma}), as interlocking is less likely to occur.

We thus find that for the general case of rough surfaces, and for the conditions here investigated, the surface morphology dominates the minimum size of the wear debris particles. This appears independent of reductions in the interfacial adhesion strength (within the explored range of values of reduced adhesion), and is determined by the junction shear strength in the full adhesion case (i.e.\ by the bulk shear strength). We do not expect the surface morphology to necessarily dominate the average size of the wear particles, where effects of interfacial adhesion are expected to emerge -- larger system sizes than those used in this set of simulations are needed to explore this question.

\vspace{\baselineskip}
\subsection{Long timescale self-affine morphology}\label{ssec:long}
Self-affine objects differ from self-similar ones by displaying anisotropic instead of isotropic scaling~\cite{barabasi1995fractal}. For a one-dimensional surface, the scaling relation is expressed as $h \left( \xi x \right) \sim \xi^{H} h \left( x \right)$~\cite{barabasi1995fractal,meakin1998fractals}, where $h(x)$ is a function describing the surface heights as a function of the spatial coordinate $x$, $\xi$ is the scaling factor, and $H$ is the Hurst (or roughness) exponent~\cite{meakin1998fractals,mandelbrot1985self}. This relation shows how the heights scale differently than the horizontal distances, with $H$ the scaling exponent. For physical surfaces, the Hurst exponent is constrained between $0$ and $1$. In the limit $H \to 1$ isotropic scaling and, thus, self-similarity are recovered. The interest in the self-affine description of surfaces lies in the fact that, for such objects, the statistics of the surface are known at any scale once the Hurst exponent is also known. This allows to gather meaningful insights at the scale that is most convenient to investigate.

\begin{figure*}
  \centering
  \includegraphics[width=\textwidth]{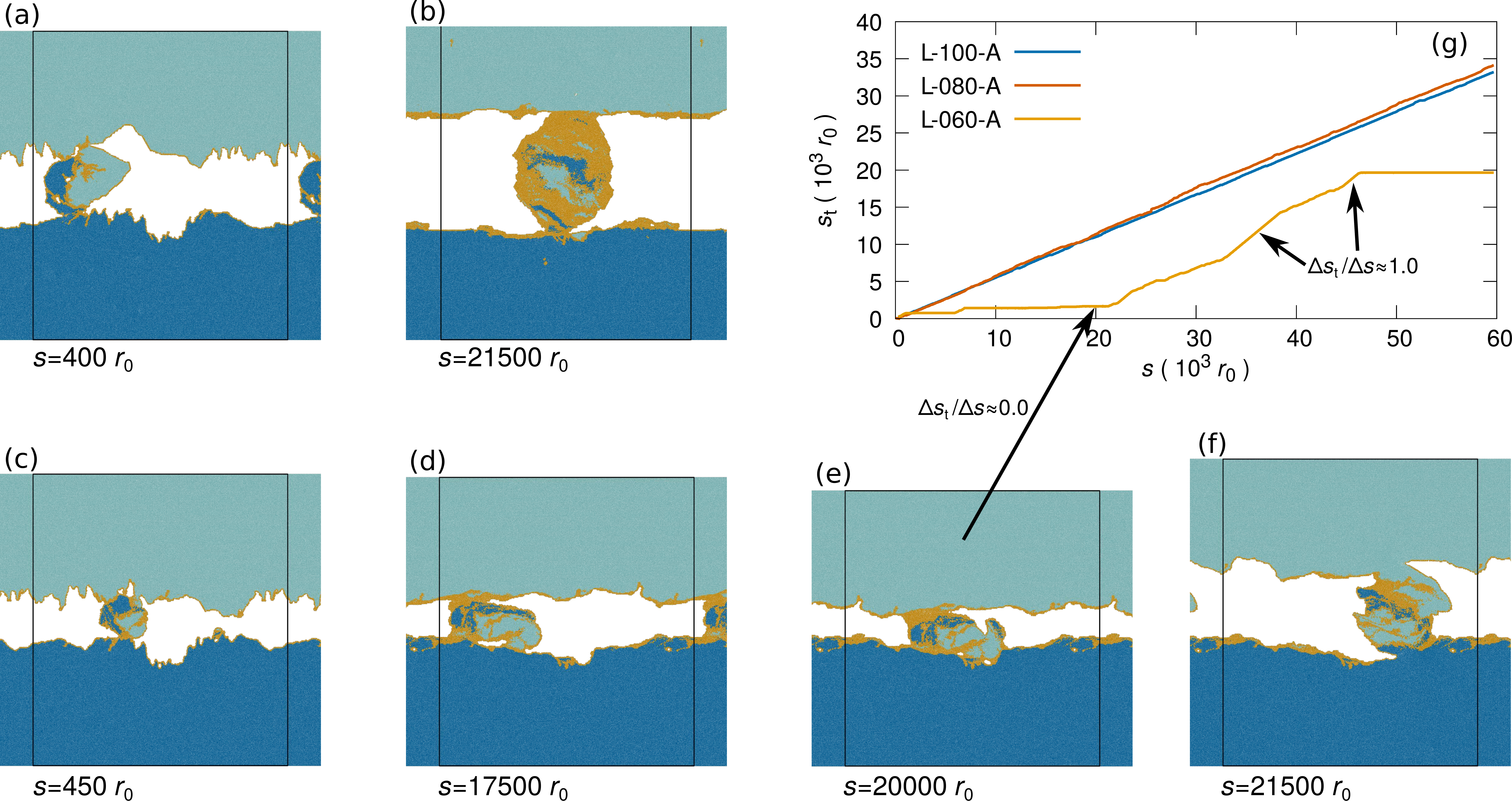}
  \caption{Long timescale evolution and effect of adhesion on debris particle motion. (a-b) Full adhesion, $\tilde{\gamma}=1.0$ (frames from full adhesion simulation L-100-A, see Supplementary Table~\ref{tab:L}). Upon sliding, contacts develop at multiple spots and grow, until cracks develop and a debris particle is formed (a). The particle then rolls between the two surfaces, wearing them and growing significantly in size (b). (c-f) Low adhesion, $\tilde{\gamma}=0.6$ (frames from reduced adhesion simulation L-060-A, see Supplementary Table~\ref{tab:L}). Starting from the same geometry of the full adhesion simulation L-100-A, upon sliding a debris particle is formed (c). This is constrained between the two surfaces and, because of the reduced interfacial adhesion and perhaps counterintuitively, it can stick for long times to one of the surfaces (e.g.\ the bottom one in (d)), the opposing surface sliding against the particle. The particle is gradually pushed into a valley (e) and, after a sticking time (where $\Delta s_\mathrm{t} / \Delta s \approx 0.0$, see arrow and (g)), it detaches again with a fracture event in a two-body like configuration (f). (g) Distance $s_\mathrm{t}$ travelled by the debris particle as a function of the sliding distance. For full and intermediate adhesion simulations (i.e.\ $\tilde{\gamma}=1.0$ and $\tilde{\gamma}=0.8$), the particle rolls most of the time. For low adhesion cases ($\tilde{\gamma}=0.6$), the particle undergoes long times of sticking to one surface (and sliding against the other). These periods are characterized by $\Delta s_\mathrm{t} / \Delta s \approx 0.0$ and $\Delta s_\mathrm{t} / \Delta s \approx 1.0$ (see arrows). In panels (a-f) colors distinguish atoms originally belonging to the top (light blue) and bottom (dark blue) surfaces. Atoms that at some previous instant were detected as surface atoms and re-assigned to the interfacial adhesion potential are depicted in yellow. In panels (a-f), black lines represent simulation box boundaries and $s$ is the sliding distance. See Supplementary Figure~\ref{fig:travel_supp} for data of $s_\mathrm{t}$ for further simulations.}
  \label{fig:H030_travel}
\end{figure*}

To investigate the effects of the reduced interfacial adhesion on the long term evolution of the surface morphology, the set of simulations L was prepared. It is characterized by $14$ long-timescale simulations -- the shortest simulated sliding distance being $60\,000$~$r_0$ and the longest $155\,650$~$r_0$. Within this set, simulations differ in the interfacial adhesion ($\tilde{\gamma} \in \{1.0,0.8,0.6\}$), the initial Hurst exponent of the surfaces ($H \in \{ 0.3, 1.0\}$), and the random seed used to generate the initial fractal surfaces. We selected such values of $H$ to avoid that the initial surfaces are already characterized by the same roughness observed at long timescales~\cite{milanese2019emergence,renard2013constant,brodsky2016constraints}. The simulations of this group are identified by a code where the first letter is 'L' for 'long timescale', the following three digits represents the value of the interfacial adhesion $\tilde{\gamma}$, and the last letter identifies a set of values $(H, \textrm{seed}, l_x)$. The simulation L-100-B, for instance, indicates a long timescale simulation, with full adhesion at the interface, and $(H, \textrm{seed}, l_x) = (0.3, 29, 339.314\textrm{~}r_0)$. The length of the investigated timescales ensures that the whole running-in phase is over and a steady-state for the roughness in terms of equivalent root mean square of heights $\sigma_\mathrm{eq}$ is reached, allowing to analyze the surface morphology in the steady-state~\cite{milanese2019emergence}. The initial geometry is then forgotten by the system and the measured morphology is a consequence of the frictional process. During each simulation, four stages are observed. Initially, the surfaces come into contact, possibly at multiple locations as the surfaces are randomly rough. The contacting spots then deform plastically, until the junction size $d$ is larger than $d^*$ and a debris particle is formed (Figures~\ref{fig:H030_travel}(a) and~\ref{fig:H030_travel}(c)). The wear particle is then constrained to roll between the surfaces, if the interfacial adhesion is large enough (Figure~\ref{fig:H030_travel}(b)). Otherwise, if the adhesion is low, the particle alternates between rolling and sticking to one surface (while sliding against the other one, Figures~\ref{fig:H030_travel}(d-f)). When the particle rolls between the two surfaces, these are continuously worn as material is transferred back and forth between the particle and each surface.

To determine if the resulting surfaces are self-affine, we investigate both their power spectral density per unit length $\Phi(q)$ and their height--height correlation function $\Delta h (\delta x)$, where $q$ and $\delta x$ are respectively the wavevector and the horizontal distance between two given points on the surface. It is known in fact that, for self-affine 1D profiles, they scale as $\Phi(q) \sim q^{-2H-1}$~\cite{mandelbrot1985self,ganti1995generalized,berry1980weierstrass} and $\Delta h (\delta x) \sim \delta x^H$~\cite{barabasi1995fractal}, respectively (see Methods for more details).

\begin{figure*}
  \centering
  \includegraphics[width=0.49\textwidth]{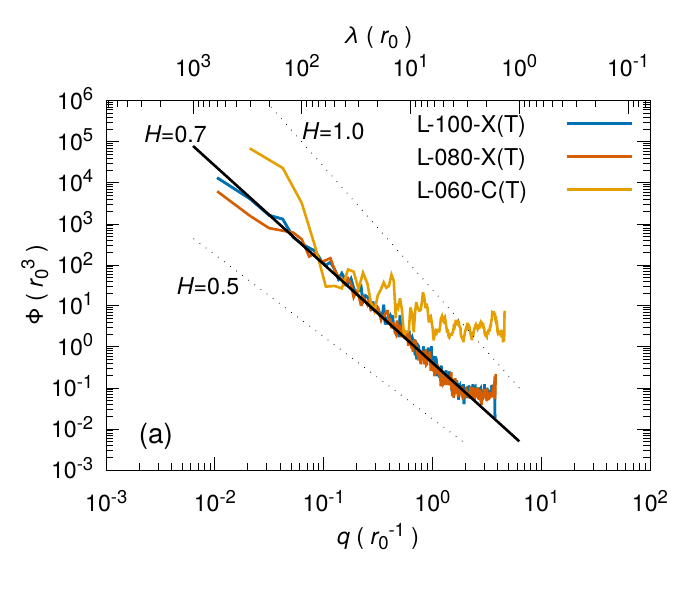}
  \includegraphics[width=0.49\textwidth]{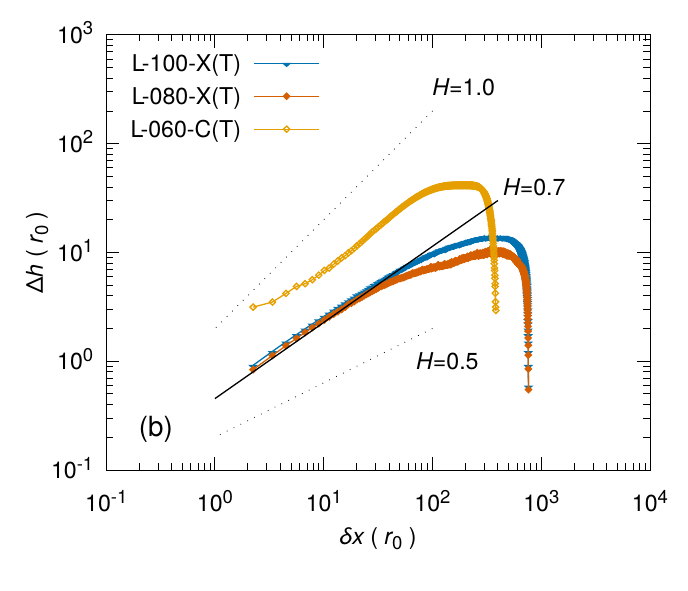}
  \caption{Surface morphology analysis. (a) PSD per unit length $\Phi$ as a function of the wavevector $q$ and the wavelength $\lambda$, where $q=2\uppi/\lambda$. (b) Height-height correlation function $\Delta h (\delta x) = \langle \left[ h(x+\delta x) - h(x) \right]^2 \rangle ^{1/2}$. The surfaces are taken from the top ('T') bodies of different simulations with different values of interfacial adhesion $\tilde{\gamma}$ and system sizes (see Supplementary Table~\ref{tab:L} for details). While top surfaces of full adhesion simulation L-100-X and low adhesion L-080-X display a self-affine morphology, this is not observed for the top surface of reduced adhesion simulation L-060-C. In this case the particle travelled a shorter distance $s_\mathrm{t}$ and did not roll for most of the simulation (see Figure~\ref{fig:H030_travel}(g)). In both panels the solid black straight guide-line corresponds to a Hurst exponent $H=0.7$. Dotted black straight guide-lines show the hypothetical slope for distributions of $H=0.5$ and $H=1.0$. Data for all the other surfaces are reported in Supplementary Figure~\ref{fig:H}.}
  \label{fig:H030b}
\end{figure*}

Figure~\ref{fig:H030b} and Supplementary Figure~\ref{fig:H} report the results of the surface analysis for the simulations in set L. The data are averaged over different independent surfaces extracted during the steady-state roughness (in terms of equivalent root mean square of heights $\sigma_\mathrm{eq}$) that follows the running-in phase~\cite{milanese2019emergence,kragelsky1981friction,rabinowicz1995friction}. While it is known that in the full adhesion case ($\tilde{\gamma}=1.0$) surfaces display a self-affine morphology characterized by a persistent Hurst exponent~\cite{milanese2019emergence}, it is observed here that the self-affine description holds also in the case of reduced interfacial adhesion ($\tilde{\gamma}<1.0$), but under some conditions.

\begin{figure*}
  \centering
  \includegraphics[width=\textwidth]{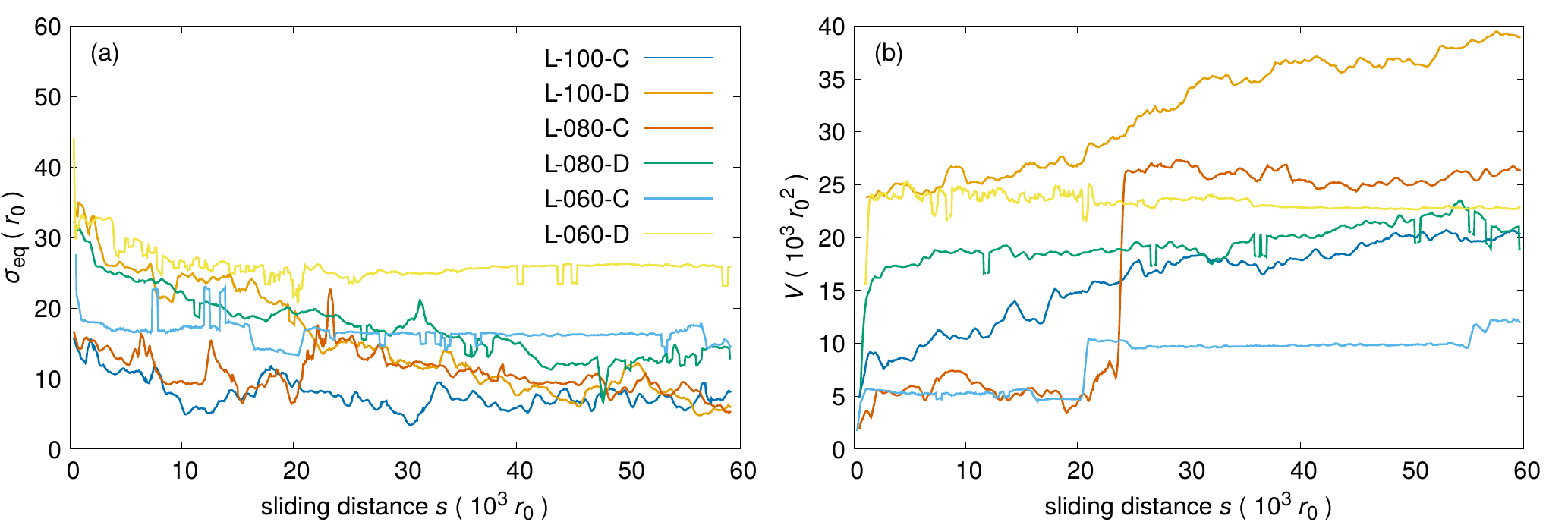}
  \caption{Evolution of the equivalent roughness $\sigma_\mathrm{eq}$ and of the wear volume $V$. (a) Evolution of $\sigma_\mathrm{eq}$ for the simulations in set L (see Supplementary Table~\ref{tab:S10}). For simulations with $\tilde{\gamma}=1.0$ and $\tilde{\gamma}=0.8$, $\sigma_\mathrm{eq}$ decreases until a steady-state is reached, where possible fluctuations due to local events can take place. For simulations with $\tilde{\gamma}=0.6$, the steady-state is not always reached (e.g.\ reduced adhesion simulation L-060-D), because of the long sticking times (see also Figure~\ref{fig:H030_travel}). (b) Evolution of the wear volume $V$ of the rolling debris particle, as defined only after its formation. The simulations with $\tilde{\gamma}=1.0$ display steady growth of the particle volume, while simulations with $\tilde{\gamma}=0.6$ are characterized by a negligible wear rate for most of the sliding distance, and the wear volume significantly increases only through fracture-like brittle events (see Figure~\ref{fig:H030_travel}). Simulations with $\tilde{\gamma}=0.8$ show an intermediate behavior. Further simulations are shown in Supplementary Figure~\ref{fig:rms_supp}}
  \label{fig:rms}
\end{figure*}

We thus investigated the distance travelled by the debris particle in each simulation (see Figure~\ref{fig:H030_travel} and Supplementary Figure~\ref{fig:travel_supp}). We observe that, in simulations for which $\tilde{\gamma}=1.0$ or $\tilde{\gamma}=0.8$, the particle travelled a comparable distance $s_\mathrm{t}$ (between $30\,000$ and $35\,000$~$r_0$) among the different simulations, and in all cases the surfaces exhibit self-affine behavior (see Figure~\ref{fig:H030b} and Supplementary Figure~\ref{fig:H}). The travelled distance is consistent with the estimation of $s_\mathrm{t} = s/2$ that is expected for a particle in perfect rolling conditions. To explain this expected value of $s_\mathrm{t}=s/2$, let us assume that both the particle and the surfaces are rigid, with the top surface sliding at constant velocity $v$ and the bottom one fixed. The highest point of the particle is then in contact with the top surface and must be sliding at velocity $v$. Similarly, the lowest point is in contact with the bottom surface and its velocity is zero. The center of the particle (which coincides with the center of mass) then rolls at velocity $v/2$, and the travelled distance is half the one of the top surface, within a given time period. When $\tilde{\gamma}=0.6$, the travelled distance $s_\mathrm{t}$ is markedly different, the value at the end of the simulations being between $20\,000$ and $26\,000$~$r_0$ (see Figure~\ref{fig:H030_travel} and Supplementary Figure~\ref{fig:travel_supp}) or over $40\,000$ (see Supplementary Figure~\ref{fig:travel_supp}). These Figures show that the particle undergoes long periods where it continuously slides against one of the surfaces (sticking to the other one), as significant portions at constant slope $\Delta s_\mathrm{t} / \Delta s = 0.0$ (sticking to bottom fixed surface) and $\Delta s_\mathrm{t} / \Delta s = 1.0$ (sticking to top sliding surface) confirm. During these periods, the particle does not roll and only works the surface against which it slides, with mechanisms that differ from the ones that take place during rolling. This is reflected by larger values of $\sigma_\mathrm{eq}$ (see Figure~\ref{fig:rms}) and the surfaces not always being characterized by a self-affine morphology (see Figure~\ref{fig:H030b}). We believe that longer sliding distances would compensate for this effect, i.e.\ the travelled distance $s_\mathrm{t}$ and the rolled distance would increase, allowing for the working of the surfaces that leads to the fractal morphology observed for larger values of the interfacial adhesion. This observation strengthens the hypothesis~\cite{milanese2019emergence} that a frictional system needs to develop third bodies that work the surfaces for them to evolve into a self-affine topography.

\vspace{\baselineskip}
\subsection{Wear rate}\label{ssec:wearrate}
The length of simulations in set L allows us to investigate also the wear rate over long distances for different values of the interfacial adhesion. Figures~\ref{fig:rms}(b) and Supplementary Figure~\ref{fig:rms_supp}(b) show the evolution of the wear volume $V$ with the sliding distance $s$. Simulations with the lowest interfacial adhesion ($\tilde{\gamma}=0.6$) display a markedly different behavior than the persistent increase in volume commonly expected, and observed for the full adhesion case~\cite{milanese2019emergence}. For reduced interfacial adhesion cases, the wear volume $V$ is characterized mostly by an almost zero wear rate, the particle volume being determined upon formation. This is due to the long sticking time. The low interfacial adhesion reduces or inhibits material transfer between the particle and the surface it slides against. And evidently no material is transferred between the particle and the surface it sticks to. Therefore the sliding motion involves only mutual deformation of the two bodies, without significant transfer of atoms (if anything, the wear particle seems to lose mass in some cases). After long periods of sticking, the deformation that takes place in the two bodies is such that the debris particle is again detached from surface it sticks to. For example, during such a period of sticking, the particle can be pushed into a valley, then forced out of it by a fracture-like event and finally it rolls for a while, until it sticks again (Figure~\ref{fig:H030_travel}(c-f)). These rare events can lead to significant local increases in the particle volume, as the detachment is fracture induced in a two-body configuration (Figure~\ref{fig:rms} and Supplementary Figure~\ref{fig:rms_supp}). The growth of the debris particle volume in the low interfacial adhesion case is then not continuous and is controlled by rare fracture events.

The case for $\tilde{\gamma}=0.8$ falls between the full and low adhesion cases. While the debris particle does not display extended sticking times, the growth rate is significantly lower than in full adhesion conditions. Rare fracture events can happen and significantly affect the wear debris volume, but they are not the predominant mechanism for mass transfer between the surfaces and the debris particle.

\section{Discussion}

In this study, we performed 2D molecular dynamics simulations of frictional systems to investigate the effect of reductions in the interfacial adhesion strength and initial surface morphology on three aspects of the adhesive wear process: the minimum size of loose wear particles, the long-term evolution of the surface roughness, and the wear rate. Our results show that, within the high adhesion regime that we explored, reducing the interfacial adhesion does not fundamentally change the nature of the processes occurring when two rough surfaces slide against one another and transition into a three-body configuration.

When the initial surfaces are self-affine, as is commonly expected in real applications, reducing the interfacial adhesion does not significantly affect the minimum initial volume of the debris particle that forms in the early stages of the sliding process. The random surface morphology thus governs the minimum size of the debris particles, which is then predicted by the critical length scale $d^*$ estimated with the values of the bulk properties (full adhesion situation). This is potentially relevant to the many engineering applications where the minimum size of wear fragments is of particular interest -- for instance in the transport industry, where particle emissions play an important role in health hazards that are linked to airborne particles.

Reduced interfacial adhesion nevertheless slows down some of the processes taking place during adhesive wear, namely the evolution of the surfaces into a self-affine morphology and the debris particle growth. Over long timescales, low values of interfacial adhesion increase the possibility of the debris particle to continuously slide against one of the surfaces (and stick to the other one), almost in a temporary two-body configuration, altering the wear mechanisms. If such periods are not too long with respect to the sliding distance, the particle still has time to roll against the surfaces and work them, and the surfaces finally exhibit a self-affine morphology, otherwise no fractal scaling is observed. Furthermore, during these periods where the particle sticks to one surface, the wear rate decreases significantly and it can become negligible.

Finally, we note that these conclusions are drawn on the basis of two-dimensional simulations, as the investigated long timescales are computationally challenging for three-dimensional systems. While analogous observations have been previously extended from 2D to 3D~\cite{aghababaei2016critical,aghababaei2017debris,brink2019adhesive}, further work is needed to extend our conclusions to three-dimensional systems and to engineering applications.


\bibliography{mybibfile}

\begin{thebibliography}{44}%
\makeatletter
\providecommand \@ifxundefined [1]{%
 \@ifx{#1\undefined}
}%
\providecommand \@ifnum [1]{%
 \ifnum #1\expandafter \@firstoftwo
 \else \expandafter \@secondoftwo
 \fi
}%
\providecommand \@ifx [1]{%
 \ifx #1\expandafter \@firstoftwo
 \else \expandafter \@secondoftwo
 \fi
}%
\providecommand \natexlab [1]{#1}%
\providecommand \enquote  [1]{``#1''}%
\providecommand \bibnamefont  [1]{#1}%
\providecommand \bibfnamefont [1]{#1}%
\providecommand \citenamefont [1]{#1}%
\providecommand \href@noop [0]{\@secondoftwo}%
\providecommand \href [0]{\begingroup \@sanitize@url \@href}%
\providecommand \@href[1]{\@@startlink{#1}\@@href}%
\providecommand \@@href[1]{\endgroup#1\@@endlink}%
\providecommand \@sanitize@url [0]{\catcode `\\12\catcode `\$12\catcode
  `\&12\catcode `\#12\catcode `\^12\catcode `\_12\catcode `\%12\relax}%
\providecommand \@@startlink[1]{}%
\providecommand \@@endlink[0]{}%
\providecommand \url  [0]{\begingroup\@sanitize@url \@url }%
\providecommand \@url [1]{\endgroup\@href {#1}{\urlprefix }}%
\providecommand \urlprefix  [0]{URL }%
\providecommand \Eprint [0]{\href }%
\providecommand \doibase [0]{https://doi.org/}%
\providecommand \selectlanguage [0]{\@gobble}%
\providecommand \bibinfo  [0]{\@secondoftwo}%
\providecommand \bibfield  [0]{\@secondoftwo}%
\providecommand \translation [1]{[#1]}%
\providecommand \BibitemOpen [0]{}%
\providecommand \bibitemStop [0]{}%
\providecommand \bibitemNoStop [0]{.\EOS\space}%
\providecommand \EOS [0]{\spacefactor3000\relax}%
\providecommand \BibitemShut  [1]{\csname bibitem#1\endcsname}%
\let\auto@bib@innerbib\@empty
\bibitem [{\citenamefont {Samet}\ \emph {et~al.}(2000)\citenamefont {Samet},
  \citenamefont {Dominici}, \citenamefont {Curriero}, \citenamefont {Coursac},\
  and\ \citenamefont {Zeger}}]{samet2000fine}%
  \BibitemOpen
  \bibfield  {author} {\bibinfo {author} {\bibfnamefont {J.~M.}\ \bibnamefont
  {Samet}}, \bibinfo {author} {\bibfnamefont {F.}~\bibnamefont {Dominici}},
  \bibinfo {author} {\bibfnamefont {F.~C.}\ \bibnamefont {Curriero}}, \bibinfo
  {author} {\bibfnamefont {I.}~\bibnamefont {Coursac}},\ and\ \bibinfo {author}
  {\bibfnamefont {S.~L.}\ \bibnamefont {Zeger}},\ }\bibfield  {title} {\bibinfo
  {title} {Fine particulate air pollution and mortality in 20 us cities,
  1987--1994},\ }\href@noop {} {\bibfield  {journal} {\bibinfo  {journal} {New
  England Journal of Medicine}\ }\textbf {\bibinfo {volume} {343}},\ \bibinfo
  {pages} {1742} (\bibinfo {year} {2000})}\BibitemShut {NoStop}%
\bibitem [{\citenamefont {Pope~III}\ \emph {et~al.}(2002)\citenamefont
  {Pope~III}, \citenamefont {Burnett}, \citenamefont {Thun}, \citenamefont
  {Calle}, \citenamefont {Krewski}, \citenamefont {Ito},\ and\ \citenamefont
  {Thurston}}]{pope2002lung}%
  \BibitemOpen
  \bibfield  {author} {\bibinfo {author} {\bibfnamefont {C.~A.}\ \bibnamefont
  {Pope~III}}, \bibinfo {author} {\bibfnamefont {R.~T.}\ \bibnamefont
  {Burnett}}, \bibinfo {author} {\bibfnamefont {M.~J.}\ \bibnamefont {Thun}},
  \bibinfo {author} {\bibfnamefont {E.~E.}\ \bibnamefont {Calle}}, \bibinfo
  {author} {\bibfnamefont {D.}~\bibnamefont {Krewski}}, \bibinfo {author}
  {\bibfnamefont {K.}~\bibnamefont {Ito}},\ and\ \bibinfo {author}
  {\bibfnamefont {G.~D.}\ \bibnamefont {Thurston}},\ }\bibfield  {title}
  {\bibinfo {title} {Lung cancer, cardiopulmonary mortality, and long-term
  exposure to fine particulate air pollution},\ }\href@noop {} {\bibfield
  {journal} {\bibinfo  {journal} {Jama}\ }\textbf {\bibinfo {volume} {287}},\
  \bibinfo {pages} {1132} (\bibinfo {year} {2002})}\BibitemShut {NoStop}%
\bibitem [{\citenamefont {Olofsson}(2011)}]{olofsson2011study}%
  \BibitemOpen
  \bibfield  {author} {\bibinfo {author} {\bibfnamefont {U.}~\bibnamefont
  {Olofsson}},\ }\bibfield  {title} {\bibinfo {title} {A study of airborne wear
  particles generated from the train traffic—block braking simulation in a
  pin-on-disc machine},\ }\href@noop {} {\bibfield  {journal} {\bibinfo
  {journal} {Wear}\ }\textbf {\bibinfo {volume} {271}},\ \bibinfo {pages} {86}
  (\bibinfo {year} {2011})}\BibitemShut {NoStop}%
\bibitem [{\citenamefont {Rabinowicz}(1995)}]{rabinowicz1995friction}%
  \BibitemOpen
  \bibfield  {author} {\bibinfo {author} {\bibfnamefont {E.}~\bibnamefont
  {Rabinowicz}},\ }\href@noop {} {\emph {\bibinfo {title} {Friction and wear of
  materials}}}\ (\bibinfo  {publisher} {Wiley, New York},\ \bibinfo {year}
  {1995})\BibitemShut {NoStop}%
\bibitem [{\citenamefont {Godet}(1984)}]{godet1984third}%
  \BibitemOpen
  \bibfield  {author} {\bibinfo {author} {\bibfnamefont {M.}~\bibnamefont
  {Godet}},\ }\bibfield  {title} {\bibinfo {title} {The third-body approach: a
  mechanical view of wear},\ }\href@noop {} {\bibfield  {journal} {\bibinfo
  {journal} {Wear}\ }\textbf {\bibinfo {volume} {100}},\ \bibinfo {pages} {437}
  (\bibinfo {year} {1984})}\BibitemShut {NoStop}%
\bibitem [{\citenamefont {Bhaskaran}\ \emph {et~al.}(2010)\citenamefont
  {Bhaskaran}, \citenamefont {Gotsmann}, \citenamefont {Sebastian},
  \citenamefont {Drechsler}, \citenamefont {Lantz}, \citenamefont {Despont},
  \citenamefont {Jaroenapibal}, \citenamefont {Carpick}, \citenamefont {Chen},\
  and\ \citenamefont {Sridharan}}]{bhaskaran2010ultralow}%
  \BibitemOpen
  \bibfield  {author} {\bibinfo {author} {\bibfnamefont {H.}~\bibnamefont
  {Bhaskaran}}, \bibinfo {author} {\bibfnamefont {B.}~\bibnamefont {Gotsmann}},
  \bibinfo {author} {\bibfnamefont {A.}~\bibnamefont {Sebastian}}, \bibinfo
  {author} {\bibfnamefont {U.}~\bibnamefont {Drechsler}}, \bibinfo {author}
  {\bibfnamefont {M.~A.}\ \bibnamefont {Lantz}}, \bibinfo {author}
  {\bibfnamefont {M.}~\bibnamefont {Despont}}, \bibinfo {author} {\bibfnamefont
  {P.}~\bibnamefont {Jaroenapibal}}, \bibinfo {author} {\bibfnamefont {R.~W.}\
  \bibnamefont {Carpick}}, \bibinfo {author} {\bibfnamefont {Y.}~\bibnamefont
  {Chen}},\ and\ \bibinfo {author} {\bibfnamefont {K.}~\bibnamefont
  {Sridharan}},\ }\bibfield  {title} {\bibinfo {title} {Ultralow nanoscale wear
  through atom-by-atom attrition in silicon-containing diamond-like carbon},\
  }\href@noop {} {\bibfield  {journal} {\bibinfo  {journal} {Nature
  Nanotechnology}\ }\textbf {\bibinfo {volume} {5}},\ \bibinfo {pages} {181}
  (\bibinfo {year} {2010})}\BibitemShut {NoStop}%
\bibitem [{\citenamefont {Schirmeisen}(2013)}]{schirmeisen2013wear}%
  \BibitemOpen
  \bibfield  {author} {\bibinfo {author} {\bibfnamefont {A.}~\bibnamefont
  {Schirmeisen}},\ }\bibfield  {title} {\bibinfo {title} {Wear: One atom after
  the other},\ }\href@noop {} {\bibfield  {journal} {\bibinfo  {journal}
  {Nature nanotechnology}\ }\textbf {\bibinfo {volume} {8}},\ \bibinfo {pages}
  {81} (\bibinfo {year} {2013})}\BibitemShut {NoStop}%
\bibitem [{\citenamefont {Jacobs}\ and\ \citenamefont
  {Carpick}(2013)}]{jacobs2013nanoscale}%
  \BibitemOpen
  \bibfield  {author} {\bibinfo {author} {\bibfnamefont {T.~D.}\ \bibnamefont
  {Jacobs}}\ and\ \bibinfo {author} {\bibfnamefont {R.~W.}\ \bibnamefont
  {Carpick}},\ }\bibfield  {title} {\bibinfo {title} {Nanoscale wear as a
  stress-assisted chemical reaction},\ }\href@noop {} {\bibfield  {journal}
  {\bibinfo  {journal} {Nature Nanotechnology}\ }\textbf {\bibinfo {volume}
  {8}},\ \bibinfo {pages} {108} (\bibinfo {year} {2013})}\BibitemShut {NoStop}%
\bibitem [{\citenamefont {Yang}\ \emph {et~al.}(2016)\citenamefont {Yang},
  \citenamefont {Huang},\ and\ \citenamefont {Shi}}]{yang2016adhesion}%
  \BibitemOpen
  \bibfield  {author} {\bibinfo {author} {\bibfnamefont {Y.}~\bibnamefont
  {Yang}}, \bibinfo {author} {\bibfnamefont {L.}~\bibnamefont {Huang}},\ and\
  \bibinfo {author} {\bibfnamefont {Y.}~\bibnamefont {Shi}},\ }\bibfield
  {title} {\bibinfo {title} {Adhesion suppresses atomic wear in single-asperity
  sliding},\ }\href@noop {} {\bibfield  {journal} {\bibinfo  {journal} {Wear}\
  }\textbf {\bibinfo {volume} {352}},\ \bibinfo {pages} {31} (\bibinfo {year}
  {2016})}\BibitemShut {NoStop}%
\bibitem [{\citenamefont {Shao}\ \emph {et~al.}(2017)\citenamefont {Shao},
  \citenamefont {Jacobs}, \citenamefont {Jiang}, \citenamefont {Turner},
  \citenamefont {Carpick},\ and\ \citenamefont {Falk}}]{shao2017multibond}%
  \BibitemOpen
  \bibfield  {author} {\bibinfo {author} {\bibfnamefont {Y.}~\bibnamefont
  {Shao}}, \bibinfo {author} {\bibfnamefont {T.~D.}\ \bibnamefont {Jacobs}},
  \bibinfo {author} {\bibfnamefont {Y.}~\bibnamefont {Jiang}}, \bibinfo
  {author} {\bibfnamefont {K.~T.}\ \bibnamefont {Turner}}, \bibinfo {author}
  {\bibfnamefont {R.~W.}\ \bibnamefont {Carpick}},\ and\ \bibinfo {author}
  {\bibfnamefont {M.~L.}\ \bibnamefont {Falk}},\ }\bibfield  {title} {\bibinfo
  {title} {Multibond model of single-asperity tribochemical wear at the
  nanoscale},\ }\href@noop {} {\bibfield  {journal} {\bibinfo  {journal} {ACS
  Applied Materials \& Interfaces}\ }\textbf {\bibinfo {volume} {9}},\ \bibinfo
  {pages} {35333} (\bibinfo {year} {2017})}\BibitemShut {NoStop}%
\bibitem [{\citenamefont {Liu}\ \emph {et~al.}(2017)\citenamefont {Liu},
  \citenamefont {Jiang}, \citenamefont {Grierson}, \citenamefont {Sridharan},
  \citenamefont {Shao}, \citenamefont {Jacobs}, \citenamefont {Falk},
  \citenamefont {Carpick},\ and\ \citenamefont
  {Turner}}]{liu2017tribochemical}%
  \BibitemOpen
  \bibfield  {author} {\bibinfo {author} {\bibfnamefont {J.}~\bibnamefont
  {Liu}}, \bibinfo {author} {\bibfnamefont {Y.}~\bibnamefont {Jiang}}, \bibinfo
  {author} {\bibfnamefont {D.~S.}\ \bibnamefont {Grierson}}, \bibinfo {author}
  {\bibfnamefont {K.}~\bibnamefont {Sridharan}}, \bibinfo {author}
  {\bibfnamefont {Y.}~\bibnamefont {Shao}}, \bibinfo {author} {\bibfnamefont
  {T.~D.}\ \bibnamefont {Jacobs}}, \bibinfo {author} {\bibfnamefont {M.~L.}\
  \bibnamefont {Falk}}, \bibinfo {author} {\bibfnamefont {R.~W.}\ \bibnamefont
  {Carpick}},\ and\ \bibinfo {author} {\bibfnamefont {K.~T.}\ \bibnamefont
  {Turner}},\ }\bibfield  {title} {\bibinfo {title} {Tribochemical wear of
  diamond-like carbon-coated atomic force microscope tips},\ }\href@noop {}
  {\bibfield  {journal} {\bibinfo  {journal} {ACS Applied Materials \&
  Interfaces}\ }\textbf {\bibinfo {volume} {9}},\ \bibinfo {pages} {35341}
  (\bibinfo {year} {2017})}\BibitemShut {NoStop}%
\bibitem [{\citenamefont {Aghababaei}\ \emph {et~al.}(2016)\citenamefont
  {Aghababaei}, \citenamefont {Warner},\ and\ \citenamefont
  {Molinari}}]{aghababaei2016critical}%
  \BibitemOpen
  \bibfield  {author} {\bibinfo {author} {\bibfnamefont {R.}~\bibnamefont
  {Aghababaei}}, \bibinfo {author} {\bibfnamefont {D.~H.}\ \bibnamefont
  {Warner}},\ and\ \bibinfo {author} {\bibfnamefont {J.-F.}\ \bibnamefont
  {Molinari}},\ }\bibfield  {title} {\bibinfo {title} {Critical length scale
  controls adhesive wear mechanisms},\ }\href@noop {} {\bibfield  {journal}
  {\bibinfo  {journal} {Nature Communications}\ }\textbf {\bibinfo {volume}
  {7}} (\bibinfo {year} {2016})}\BibitemShut {NoStop}%
\bibitem [{\citenamefont {Holm}(1946)}]{holm1946}%
  \BibitemOpen
  \bibfield  {author} {\bibinfo {author} {\bibfnamefont {R.}~\bibnamefont
  {Holm}},\ }\href@noop {} {\emph {\bibinfo {title} {Electric contacts}}}\
  (\bibinfo  {publisher} {Almqvist and Wiksells, Stockholm},\ \bibinfo {year}
  {1946})\BibitemShut {NoStop}%
\bibitem [{\citenamefont {Merkle}\ and\ \citenamefont
  {Marks}(2008)}]{merkle2008liquid}%
  \BibitemOpen
  \bibfield  {author} {\bibinfo {author} {\bibfnamefont {A.~P.}\ \bibnamefont
  {Merkle}}\ and\ \bibinfo {author} {\bibfnamefont {L.~D.}\ \bibnamefont
  {Marks}},\ }\bibfield  {title} {\bibinfo {title} {Liquid-like tribology of
  gold studied by in situ tem},\ }\href@noop {} {\bibfield  {journal} {\bibinfo
   {journal} {Wear}\ }\textbf {\bibinfo {volume} {265}},\ \bibinfo {pages}
  {1864} (\bibinfo {year} {2008})}\BibitemShut {NoStop}%
\bibitem [{\citenamefont {Milanese}\ \emph {et~al.}(2019)\citenamefont
  {Milanese}, \citenamefont {Brink}, \citenamefont {Aghababaei},\ and\
  \citenamefont {Molinari}}]{milanese2019emergence}%
  \BibitemOpen
  \bibfield  {author} {\bibinfo {author} {\bibfnamefont {E.}~\bibnamefont
  {Milanese}}, \bibinfo {author} {\bibfnamefont {T.}~\bibnamefont {Brink}},
  \bibinfo {author} {\bibfnamefont {R.}~\bibnamefont {Aghababaei}},\ and\
  \bibinfo {author} {\bibfnamefont {J.-F.}\ \bibnamefont {Molinari}},\
  }\bibfield  {title} {\bibinfo {title} {Emergence of self-affine surfaces
  during adhesive wear},\ }\href@noop {} {\bibfield  {journal} {\bibinfo
  {journal} {Nature Communications}\ }\textbf {\bibinfo {volume} {10}},\
  \bibinfo {pages} {1116} (\bibinfo {year} {2019})}\BibitemShut {NoStop}%
\bibitem [{\citenamefont {Archard}(1953)}]{archard1953contact}%
  \BibitemOpen
  \bibfield  {author} {\bibinfo {author} {\bibfnamefont {J.}~\bibnamefont
  {Archard}},\ }\bibfield  {title} {\bibinfo {title} {Contact and rubbing of
  flat surfaces},\ }\href@noop {} {\bibfield  {journal} {\bibinfo  {journal}
  {Journal of Applied Physics}\ }\textbf {\bibinfo {volume} {24}},\ \bibinfo
  {pages} {981} (\bibinfo {year} {1953})}\BibitemShut {NoStop}%
\bibitem [{\citenamefont {Liu}\ \emph {et~al.}(2010{\natexlab{a}})\citenamefont
  {Liu}, \citenamefont {Notbohm}, \citenamefont {Carpick},\ and\ \citenamefont
  {Turner}}]{liu2010method}%
  \BibitemOpen
  \bibfield  {author} {\bibinfo {author} {\bibfnamefont {J.}~\bibnamefont
  {Liu}}, \bibinfo {author} {\bibfnamefont {J.~K.}\ \bibnamefont {Notbohm}},
  \bibinfo {author} {\bibfnamefont {R.~W.}\ \bibnamefont {Carpick}},\ and\
  \bibinfo {author} {\bibfnamefont {K.~T.}\ \bibnamefont {Turner}},\ }\bibfield
   {title} {\bibinfo {title} {Method for characterizing nanoscale wear of
  atomic force microscope tips},\ }\href@noop {} {\bibfield  {journal}
  {\bibinfo  {journal} {ACS Nano}\ }\textbf {\bibinfo {volume} {4}},\ \bibinfo
  {pages} {3763} (\bibinfo {year} {2010}{\natexlab{a}})}\BibitemShut {NoStop}%
\bibitem [{\citenamefont {Liu}\ \emph {et~al.}(2010{\natexlab{b}})\citenamefont
  {Liu}, \citenamefont {Grierson}, \citenamefont {Moldovan}, \citenamefont
  {Notbohm}, \citenamefont {Li}, \citenamefont {Jaroenapibal}, \citenamefont
  {O'Connor}, \citenamefont {Sumant}, \citenamefont {Neelakantan},
  \citenamefont {Carlisle} \emph {et~al.}}]{liu2010preventing}%
  \BibitemOpen
  \bibfield  {author} {\bibinfo {author} {\bibfnamefont {J.}~\bibnamefont
  {Liu}}, \bibinfo {author} {\bibfnamefont {D.~S.}\ \bibnamefont {Grierson}},
  \bibinfo {author} {\bibfnamefont {N.}~\bibnamefont {Moldovan}}, \bibinfo
  {author} {\bibfnamefont {J.}~\bibnamefont {Notbohm}}, \bibinfo {author}
  {\bibfnamefont {S.}~\bibnamefont {Li}}, \bibinfo {author} {\bibfnamefont
  {P.}~\bibnamefont {Jaroenapibal}}, \bibinfo {author} {\bibfnamefont
  {S.}~\bibnamefont {O'Connor}}, \bibinfo {author} {\bibfnamefont
  {A.}~\bibnamefont {Sumant}}, \bibinfo {author} {\bibfnamefont
  {N.}~\bibnamefont {Neelakantan}}, \bibinfo {author} {\bibfnamefont {J.~A.}\
  \bibnamefont {Carlisle}}, \emph {et~al.},\ }\bibfield  {title} {\bibinfo
  {title} {Preventing nanoscale wear of atomic force microscopy tips through
  the use of monolithic ultrananocrystalline diamond probes},\ }\href@noop {}
  {\bibfield  {journal} {\bibinfo  {journal} {Small}\ }\textbf {\bibinfo
  {volume} {6}},\ \bibinfo {pages} {1140} (\bibinfo {year}
  {2010}{\natexlab{b}})}\BibitemShut {NoStop}%
\bibitem [{\citenamefont {Aghababaei}\ \emph {et~al.}(2017)\citenamefont
  {Aghababaei}, \citenamefont {Warner},\ and\ \citenamefont
  {Molinari}}]{aghababaei2017debris}%
  \BibitemOpen
  \bibfield  {author} {\bibinfo {author} {\bibfnamefont {R.}~\bibnamefont
  {Aghababaei}}, \bibinfo {author} {\bibfnamefont {D.~H.}\ \bibnamefont
  {Warner}},\ and\ \bibinfo {author} {\bibfnamefont {J.-F.}\ \bibnamefont
  {Molinari}},\ }\bibfield  {title} {\bibinfo {title} {On the debris-level
  origins of adhesive wear},\ }\href@noop {} {\bibfield  {journal} {\bibinfo
  {journal} {Proceedings of the National Academy of Sciences}\ }\textbf
  {\bibinfo {volume} {114}},\ \bibinfo {pages} {7935} (\bibinfo {year}
  {2017})}\BibitemShut {NoStop}%
\bibitem [{\citenamefont {Aghababaei}\ \emph {et~al.}(2018)\citenamefont
  {Aghababaei}, \citenamefont {Brink},\ and\ \citenamefont
  {Molinari}}]{aghababaei2018asperity}%
  \BibitemOpen
  \bibfield  {author} {\bibinfo {author} {\bibfnamefont {R.}~\bibnamefont
  {Aghababaei}}, \bibinfo {author} {\bibfnamefont {T.}~\bibnamefont {Brink}},\
  and\ \bibinfo {author} {\bibfnamefont {J.-F.}\ \bibnamefont {Molinari}},\
  }\bibfield  {title} {\bibinfo {title} {Asperity-level origins of transition
  from mild to severe wear},\ }\href@noop {} {\bibfield  {journal} {\bibinfo
  {journal} {Physical Review Letters}\ }\textbf {\bibinfo {volume} {120}},\
  \bibinfo {pages} {186105} (\bibinfo {year} {2018})}\BibitemShut {NoStop}%
\bibitem [{\citenamefont {Fr{\'{e}}rot}\ \emph {et~al.}(2018)\citenamefont
  {Fr{\'{e}}rot}, \citenamefont {Aghababaei},\ and\ \citenamefont
  {Molinari}}]{Frerot2018}%
  \BibitemOpen
  \bibfield  {author} {\bibinfo {author} {\bibfnamefont {L.}~\bibnamefont
  {Fr{\'{e}}rot}}, \bibinfo {author} {\bibfnamefont {R.}~\bibnamefont
  {Aghababaei}},\ and\ \bibinfo {author} {\bibfnamefont {J.-F.}\ \bibnamefont
  {Molinari}},\ }\bibfield  {title} {\bibinfo {title} {A mechanistic
  understanding of the wear coefficient: From single to multiple asperities
  contact},\ }\href {https://doi.org/10.1016/j.jmps.2018.02.015} {\bibfield
  {journal} {\bibinfo  {journal} {Journal of the Mechanics and Physics of
  Solids}\ }\textbf {\bibinfo {volume} {114}},\ \bibinfo {pages} {172}
  (\bibinfo {year} {2018})}\BibitemShut {NoStop}%
\bibitem [{\citenamefont {Pham-Ba}\ \emph {et~al.}(2019)\citenamefont
  {Pham-Ba}, \citenamefont {Brink},\ and\ \citenamefont
  {Molinari}}]{pham2019adhesive}%
  \BibitemOpen
  \bibfield  {author} {\bibinfo {author} {\bibfnamefont {S.}~\bibnamefont
  {Pham-Ba}}, \bibinfo {author} {\bibfnamefont {T.}~\bibnamefont {Brink}},\
  and\ \bibinfo {author} {\bibfnamefont {J.-F.}\ \bibnamefont {Molinari}},\
  }\bibfield  {title} {\bibinfo {title} {Adhesive wear and interaction of
  tangentially loaded micro-contacts},\ }\href@noop {} {\bibfield  {journal}
  {\bibinfo  {journal} {International Journal of Solids and Structures}\ }
  (\bibinfo {year} {2019})}\BibitemShut {NoStop}%
\bibitem [{\citenamefont {Brink}\ and\ \citenamefont
  {Molinari}(2019)}]{brink2019adhesive}%
  \BibitemOpen
  \bibfield  {author} {\bibinfo {author} {\bibfnamefont {T.}~\bibnamefont
  {Brink}}\ and\ \bibinfo {author} {\bibfnamefont {J.-F.}\ \bibnamefont
  {Molinari}},\ }\bibfield  {title} {\bibinfo {title} {Adhesive wear mechanisms
  in the presence of weak interfaces: Insights from an amorphous model
  system},\ }\href@noop {} {\bibfield  {journal} {\bibinfo  {journal} {Physical
  Review Materials}\ }\textbf {\bibinfo {volume} {3}},\ \bibinfo {pages}
  {053604} (\bibinfo {year} {2019})}\BibitemShut {NoStop}%
\bibitem [{\citenamefont {Aghababaei}(2019)}]{aghababaei2019effect}%
  \BibitemOpen
  \bibfield  {author} {\bibinfo {author} {\bibfnamefont {R.}~\bibnamefont
  {Aghababaei}},\ }\bibfield  {title} {\bibinfo {title} {Effect of adhesion on
  material removal during adhesive wear},\ }\href@noop {} {\bibfield  {journal}
  {\bibinfo  {journal} {Physical Review Materials}\ }\textbf {\bibinfo {volume}
  {3}},\ \bibinfo {pages} {063604} (\bibinfo {year} {2019})}\BibitemShut
  {NoStop}%
\bibitem [{\citenamefont {Renard}\ \emph {et~al.}(2013)\citenamefont {Renard},
  \citenamefont {Candela},\ and\ \citenamefont
  {Bouchaud}}]{renard2013constant}%
  \BibitemOpen
  \bibfield  {author} {\bibinfo {author} {\bibfnamefont {F.}~\bibnamefont
  {Renard}}, \bibinfo {author} {\bibfnamefont {T.}~\bibnamefont {Candela}},\
  and\ \bibinfo {author} {\bibfnamefont {E.}~\bibnamefont {Bouchaud}},\
  }\bibfield  {title} {\bibinfo {title} {Constant dimensionality of fault
  roughness from the scale of micro-fractures to the scale of continents},\
  }\href@noop {} {\bibfield  {journal} {\bibinfo  {journal} {Geophysical
  Research Letters}\ }\textbf {\bibinfo {volume} {40}},\ \bibinfo {pages} {83}
  (\bibinfo {year} {2013})}\BibitemShut {NoStop}%
\bibitem [{\citenamefont {Sayles}\ and\ \citenamefont
  {Thomas}(1978)}]{sayles1978surface}%
  \BibitemOpen
  \bibfield  {author} {\bibinfo {author} {\bibfnamefont {R.~S.}\ \bibnamefont
  {Sayles}}\ and\ \bibinfo {author} {\bibfnamefont {T.~R.}\ \bibnamefont
  {Thomas}},\ }\bibfield  {title} {\bibinfo {title} {Surface topography as a
  nonstationary random process},\ }\href@noop {} {\bibfield  {journal}
  {\bibinfo  {journal} {Nature}\ }\textbf {\bibinfo {volume} {271}},\ \bibinfo
  {pages} {431} (\bibinfo {year} {1978})}\BibitemShut {NoStop}%
\bibitem [{\citenamefont {Majumdar}\ and\ \citenamefont
  {Tien}(1990)}]{majumdar1990fractal}%
  \BibitemOpen
  \bibfield  {author} {\bibinfo {author} {\bibfnamefont {A.}~\bibnamefont
  {Majumdar}}\ and\ \bibinfo {author} {\bibfnamefont {C.}~\bibnamefont
  {Tien}},\ }\bibfield  {title} {\bibinfo {title} {Fractal characterization and
  simulation of rough surfaces},\ }\href@noop {} {\bibfield  {journal}
  {\bibinfo  {journal} {Wear}\ }\textbf {\bibinfo {volume} {136}},\ \bibinfo
  {pages} {313} (\bibinfo {year} {1990})}\BibitemShut {NoStop}%
\bibitem [{\citenamefont {Persson}\ \emph {et~al.}(2004)\citenamefont
  {Persson}, \citenamefont {Albohr}, \citenamefont {Tartaglino}, \citenamefont
  {Volokitin},\ and\ \citenamefont {Tosatti}}]{persson2004nature}%
  \BibitemOpen
  \bibfield  {author} {\bibinfo {author} {\bibfnamefont {B.}~\bibnamefont
  {Persson}}, \bibinfo {author} {\bibfnamefont {O.}~\bibnamefont {Albohr}},
  \bibinfo {author} {\bibfnamefont {U.}~\bibnamefont {Tartaglino}}, \bibinfo
  {author} {\bibfnamefont {A.}~\bibnamefont {Volokitin}},\ and\ \bibinfo
  {author} {\bibfnamefont {E.}~\bibnamefont {Tosatti}},\ }\bibfield  {title}
  {\bibinfo {title} {On the nature of surface roughness with application to
  contact mechanics, sealing, rubber friction and adhesion},\ }\href@noop {}
  {\bibfield  {journal} {\bibinfo  {journal} {Journal of Physics: Condensed
  Matter}\ }\textbf {\bibinfo {volume} {17}},\ \bibinfo {pages} {R1} (\bibinfo
  {year} {2004})}\BibitemShut {NoStop}%
\bibitem [{\citenamefont {Morse}(1929)}]{morse1929diatomic}%
  \BibitemOpen
  \bibfield  {author} {\bibinfo {author} {\bibfnamefont {P.~M.}\ \bibnamefont
  {Morse}},\ }\bibfield  {title} {\bibinfo {title} {Diatomic molecules
  according to the wave mechanics. {II}. {V}ibrational levels},\ }\href@noop {}
  {\bibfield  {journal} {\bibinfo  {journal} {Physical Review}\ }\textbf
  {\bibinfo {volume} {34}},\ \bibinfo {pages} {57} (\bibinfo {year}
  {1929})}\BibitemShut {NoStop}%
\bibitem [{\citenamefont {Plimpton}(1995)}]{plimpton1995fast}%
  \BibitemOpen
  \bibfield  {author} {\bibinfo {author} {\bibfnamefont {S.}~\bibnamefont
  {Plimpton}},\ }\bibfield  {title} {\bibinfo {title} {Fast parallel algorithms
  for short-range molecular dynamics},\ }\href@noop {} {\bibfield  {journal}
  {\bibinfo  {journal} {Journal of Computational Physics}\ }\textbf {\bibinfo
  {volume} {117}},\ \bibinfo {pages} {1} (\bibinfo {year} {1995})}\BibitemShut
  {NoStop}%
\bibitem [{\citenamefont {Jacobs}\ \emph {et~al.}(2017)\citenamefont {Jacobs},
  \citenamefont {Junge},\ and\ \citenamefont
  {Pastewka}}]{jacobs2017quantitative}%
  \BibitemOpen
  \bibfield  {author} {\bibinfo {author} {\bibfnamefont {T.~D.}\ \bibnamefont
  {Jacobs}}, \bibinfo {author} {\bibfnamefont {T.}~\bibnamefont {Junge}},\ and\
  \bibinfo {author} {\bibfnamefont {L.}~\bibnamefont {Pastewka}},\ }\bibfield
  {title} {\bibinfo {title} {Quantitative characterization of surface
  topography using spectral analysis},\ }\href@noop {} {\bibfield  {journal}
  {\bibinfo  {journal} {Surface Topography: Metrology and Properties}\ }\textbf
  {\bibinfo {volume} {5}},\ \bibinfo {pages} {013001} (\bibinfo {year}
  {2017})}\BibitemShut {NoStop}%
\bibitem [{\citenamefont {Barab{\'a}si}\ and\ \citenamefont
  {Stanley}(1995)}]{barabasi1995fractal}%
  \BibitemOpen
  \bibfield  {author} {\bibinfo {author} {\bibfnamefont {A.-L.}\ \bibnamefont
  {Barab{\'a}si}}\ and\ \bibinfo {author} {\bibfnamefont {H.~E.}\ \bibnamefont
  {Stanley}},\ }\href@noop {} {\emph {\bibinfo {title} {Fractal concepts in
  surface growth}}}\ (\bibinfo  {publisher} {Cambridge university press},\
  \bibinfo {year} {1995})\BibitemShut {NoStop}%
\bibitem [{\citenamefont {Meakin}(1998)}]{meakin1998fractals}%
  \BibitemOpen
  \bibfield  {author} {\bibinfo {author} {\bibfnamefont {P.}~\bibnamefont
  {Meakin}},\ }\href@noop {} {\emph {\bibinfo {title} {Fractals, scaling and
  growth far from equilibrium}}},\ Vol.~\bibinfo {volume} {5}\ (\bibinfo
  {publisher} {Cambridge university press},\ \bibinfo {year}
  {1998})\BibitemShut {NoStop}%
\bibitem [{\citenamefont {Mandelbrot}(1985)}]{mandelbrot1985self}%
  \BibitemOpen
  \bibfield  {author} {\bibinfo {author} {\bibfnamefont {B.~B.}\ \bibnamefont
  {Mandelbrot}},\ }\bibfield  {title} {\bibinfo {title} {Self-affine fractals
  and fractal dimension},\ }\href@noop {} {\bibfield  {journal} {\bibinfo
  {journal} {Physica Scripta}\ }\textbf {\bibinfo {volume} {32}},\ \bibinfo
  {pages} {257} (\bibinfo {year} {1985})}\BibitemShut {NoStop}%
\bibitem [{\citenamefont {Ganti}\ and\ \citenamefont
  {Bhushan}(1995)}]{ganti1995generalized}%
  \BibitemOpen
  \bibfield  {author} {\bibinfo {author} {\bibfnamefont {S.}~\bibnamefont
  {Ganti}}\ and\ \bibinfo {author} {\bibfnamefont {B.}~\bibnamefont
  {Bhushan}},\ }\bibfield  {title} {\bibinfo {title} {Generalized fractal
  analysis and its applications to engineering surfaces},\ }\href@noop {}
  {\bibfield  {journal} {\bibinfo  {journal} {Wear}\ }\textbf {\bibinfo
  {volume} {180}},\ \bibinfo {pages} {17} (\bibinfo {year} {1995})}\BibitemShut
  {NoStop}%
\bibitem [{\citenamefont {Berry}\ and\ \citenamefont
  {Lewis}(1980)}]{berry1980weierstrass}%
  \BibitemOpen
  \bibfield  {author} {\bibinfo {author} {\bibfnamefont {M.}~\bibnamefont
  {Berry}}\ and\ \bibinfo {author} {\bibfnamefont {Z.}~\bibnamefont {Lewis}},\
  }\bibfield  {title} {\bibinfo {title} {On the {W}eierstrass-{M}andelbrot
  fractal function},\ }in\ \href@noop {} {\emph {\bibinfo {booktitle}
  {Proceedings of the Royal Society of London A: Mathematical, Physical and
  Engineering Sciences}}},\ Vol.\ \bibinfo {volume} {370}\ (\bibinfo
  {organization} {The Royal Society},\ \bibinfo {year} {1980})\ pp.\ \bibinfo
  {pages} {459--484}\BibitemShut {NoStop}%
\bibitem [{\citenamefont {Press}\ \emph {et~al.}(2007)\citenamefont {Press},
  \citenamefont {Teukolsky}, \citenamefont {Vetterling},\ and\ \citenamefont
  {Flannery}}]{press2007numerical}%
  \BibitemOpen
  \bibfield  {author} {\bibinfo {author} {\bibfnamefont {W.~H.}\ \bibnamefont
  {Press}}, \bibinfo {author} {\bibfnamefont {S.~A.}\ \bibnamefont
  {Teukolsky}}, \bibinfo {author} {\bibfnamefont {W.~T.}\ \bibnamefont
  {Vetterling}},\ and\ \bibinfo {author} {\bibfnamefont {B.~P.}\ \bibnamefont
  {Flannery}},\ }\href@noop {} {\emph {\bibinfo {title} {Numerical recipes 3rd
  edition: The art of scientific computing}}}\ (\bibinfo  {publisher}
  {Cambridge university press},\ \bibinfo {year} {2007})\BibitemShut {NoStop}%
\bibitem [{\citenamefont {VanderPlas}(2018)}]{vanderplas2018understanding}%
  \BibitemOpen
  \bibfield  {author} {\bibinfo {author} {\bibfnamefont {J.~T.}\ \bibnamefont
  {VanderPlas}},\ }\bibfield  {title} {\bibinfo {title} {Understanding the
  {L}omb--{S}cargle periodogram},\ }\href@noop {} {\bibfield  {journal}
  {\bibinfo  {journal} {The Astrophysical Journal Supplement Series}\ }\textbf
  {\bibinfo {volume} {236}},\ \bibinfo {pages} {16} (\bibinfo {year}
  {2018})}\BibitemShut {NoStop}%
\bibitem [{\citenamefont {Bhushan}(2000)}]{bhushan2000modern}%
  \BibitemOpen
  \bibfield  {author} {\bibinfo {author} {\bibfnamefont {B.}~\bibnamefont
  {Bhushan}},\ }\bibfield  {title} {\bibinfo {title} {Surface roughness
  analysis and measurement techniques},\ }in\ \href@noop {} {\emph {\bibinfo
  {booktitle} {Modern tribology handbook, two volume set}}},\ \bibinfo {editor}
  {edited by\ \bibinfo {editor} {\bibfnamefont {B.}~\bibnamefont {Bhushan}}}\
  (\bibinfo  {publisher} {CRC press},\ \bibinfo {year} {2000})\ Chap.~\bibinfo
  {chapter} {4}, pp.\ \bibinfo {pages} {49--120}\BibitemShut {NoStop}%
\bibitem [{\citenamefont {Stukowski}(2009)}]{stukowski2009visualization}%
  \BibitemOpen
  \bibfield  {author} {\bibinfo {author} {\bibfnamefont {A.}~\bibnamefont
  {Stukowski}},\ }\bibfield  {title} {\bibinfo {title} {Visualization and
  analysis of atomistic simulation data with {OVITO}--the {O}pen
  {V}isualization {T}ool},\ }\href@noop {} {\bibfield  {journal} {\bibinfo
  {journal} {Modelling and Simulation in Materials Science and Engineering}\
  }\textbf {\bibinfo {volume} {18}},\ \bibinfo {pages} {015012} (\bibinfo
  {year} {2009})}\BibitemShut {NoStop}%
\bibitem [{\citenamefont {Kirkpatrick}\ \emph {et~al.}(1983)\citenamefont
  {Kirkpatrick}, \citenamefont {Gelatt},\ and\ \citenamefont
  {Vecchi}}]{Kirkpatrick1983}%
  \BibitemOpen
  \bibfield  {author} {\bibinfo {author} {\bibfnamefont {S.}~\bibnamefont
  {Kirkpatrick}}, \bibinfo {author} {\bibfnamefont {C.~D.}\ \bibnamefont
  {Gelatt}},\ and\ \bibinfo {author} {\bibfnamefont {M.~P.}\ \bibnamefont
  {Vecchi}},\ }\bibfield  {title} {\bibinfo {title} {Optimization by simulated
  annealing},\ }\href {https://doi.org/10.1126/science.220.4598.671} {\bibfield
   {journal} {\bibinfo  {journal} {Science}\ }\textbf {\bibinfo {volume}
  {220}},\ \bibinfo {pages} {671} (\bibinfo {year} {1983})}\BibitemShut
  {NoStop}%
\bibitem [{\citenamefont {Metropolis}\ \emph {et~al.}(1953)\citenamefont
  {Metropolis}, \citenamefont {Rosenbluth}, \citenamefont {Rosenbluth},
  \citenamefont {Teller},\ and\ \citenamefont {Teller}}]{Metropolis1953}%
  \BibitemOpen
  \bibfield  {author} {\bibinfo {author} {\bibfnamefont {N.}~\bibnamefont
  {Metropolis}}, \bibinfo {author} {\bibfnamefont {A.~W.}\ \bibnamefont
  {Rosenbluth}}, \bibinfo {author} {\bibfnamefont {M.~N.}\ \bibnamefont
  {Rosenbluth}}, \bibinfo {author} {\bibfnamefont {A.~H.}\ \bibnamefont
  {Teller}},\ and\ \bibinfo {author} {\bibfnamefont {E.}~\bibnamefont
  {Teller}},\ }\bibfield  {title} {\bibinfo {title} {Equation of state
  calculations by fast computing machines},\ }\href
  {https://doi.org/10.1063/1.1699114} {\bibfield  {journal} {\bibinfo
  {journal} {The Journal of Chemical Physics}\ }\textbf {\bibinfo {volume}
  {21}},\ \bibinfo {pages} {1087} (\bibinfo {year} {1953})}\BibitemShut
  {NoStop}%
\bibitem [{\citenamefont {Brodsky}\ \emph {et~al.}(2016)\citenamefont
  {Brodsky}, \citenamefont {Kirkpatrick},\ and\ \citenamefont
  {Candela}}]{brodsky2016constraints}%
  \BibitemOpen
  \bibfield  {author} {\bibinfo {author} {\bibfnamefont {E.~E.}\ \bibnamefont
  {Brodsky}}, \bibinfo {author} {\bibfnamefont {J.~D.}\ \bibnamefont
  {Kirkpatrick}},\ and\ \bibinfo {author} {\bibfnamefont {T.}~\bibnamefont
  {Candela}},\ }\bibfield  {title} {\bibinfo {title} {Constraints from fault
  roughness on the scale-dependent strength of rocks},\ }\href@noop {}
  {\bibfield  {journal} {\bibinfo  {journal} {Geology}\ }\textbf {\bibinfo
  {volume} {44}},\ \bibinfo {pages} {19} (\bibinfo {year} {2016})}\BibitemShut
  {NoStop}%
\bibitem [{\citenamefont {Kragelsky}\ \emph {et~al.}(1981)\citenamefont
  {Kragelsky}, \citenamefont {Dobychin},\ and\ \citenamefont
  {Kombalov}}]{kragelsky1981friction}%
  \BibitemOpen
  \bibfield  {author} {\bibinfo {author} {\bibfnamefont {I.~V.}\ \bibnamefont
  {Kragelsky}}, \bibinfo {author} {\bibfnamefont {M.~N.}\ \bibnamefont
  {Dobychin}},\ and\ \bibinfo {author} {\bibfnamefont {V.~S.}\ \bibnamefont
  {Kombalov}},\ }\href@noop {} {\emph {\bibinfo {title} {Friction and wear:
  calculation methods}}}\ (\bibinfo  {publisher} {Pergamon Press},\ \bibinfo
  {year} {1981})\BibitemShut {NoStop}%
\end{thebibliography}%

\section*{Supplemental Material}
See Supplemental Material at (URL by publisher) for supplementary methods, figures, and tables referenced in the main text.

\section*{Acknowledgements}
This work was supported by EPFL through the use of the facilities of its Scientific IT and Application Support Center.

\section*{Authors Contributions}
 E. M. performed the numerical simulations and the analyses. R. A. developed the algorithm to detect surface atoms and re-assign the interfacial adhesion potential on-the-fly. E. M., T. B., and J.-F. M. planned the simulations sets and the wrote the manuscript. All authors participated in the discussion.

\clearpage
\newpage

\onecolumngrid
\section*{Supplementary Materials}

\subsection{Supplementary methods}

\subsubsection*{Geometrical factor $\Lambda$ for flat contacts}

In the original formulation of Ref.~\onlinecite{aghababaei2016critical}, the critical length scale $d^*$ for the ductile-to-brittle transition stems from a Griffith-like criterion applied to asperities of a generic shape. According to the criterion, if the elastic energy $E_\mathrm{el}$ is larger or equal than the adhesive energy $E_\mathrm{ad}$, crack propagation is favoured over plastic deformation, and a debris particle is formed. As the two energies scale as $E_\mathrm{el} \sim d^3$ and $E_\mathrm{ad} \sim d^2$, the minimum contact junction size at which crack propagation takes place is found for $E_\mathrm{el} = E_\mathrm{ad}$ and it has the form

\begin{align}
  d^* = \Lambda {} \frac {\Delta w G}{\tau_\mathrm{j}^2} \textrm{ ,}
  \label{eq:dstar_ramin}
\end{align}
where $\tau_\mathrm{j}$ is the junction shear strength (and in the general case depends on $\tilde{\gamma}$), $G$ is the shear modulus of the material, $\Delta w$ is twice the fracture energy and $\Lambda$ is a geometrical factor. In the derivation of Eq.~\ref{eq:dstar_ramin}, it is assumed that the elastic energy is fully contained in the asperity volume, which is uniformly loaded at the shear junction strength $\tau_\mathrm{j}$. $E_\mathrm{el}$ thus depends on both the characteristic size $d$ and the shape of the asperity. $E_\mathrm{ad}$ also depends on both, as the shape of the asperity defines the crack path and, consequently, the free surfaces that needs to be created in the fracture process. The geometrical factor $\Lambda$ takes into account the shape of the asperities and the actual stress distribution (due to this shape). The meaning of $\Lambda$ is clear in the expression of $d^*$ if Eq.~\ref{eq:dstar_ramin} is rewritten in the form used in the main text (Eq.~\ref{eq:dstar})

\begin{align}
  d^* = \Lambda {} \frac{w}{\tau_\mathrm{j}^2 / 2G} \textrm{,}
\end{align}
where $\Lambda$ regulates the ratio between the fracture energy $w=\Delta w /2$ and the stored elastic energy density $\tau_\mathrm{j}^2 / 2G$. In Ref.~\onlinecite{aghababaei2016critical}, data from molecular dynamics simulations shows that $\Lambda=1.50$ for well-defined semicircular asperities in 2D.

More recently, another formulation for $d^*$ in two dimensions was derived based on an analytical approach in Ref.~\onlinecite{pham2019adhesive}. Here, the asperity is replaced by a distributed constant shear load of magnitude $\tau_\mathrm{j}$ applied along a length $d$ on the flat surface of a semi-infinite body. An analytical expression for $E_\mathrm{el}$ in the whole body is then provided. The value of $E_\mathrm{el}$ then depends on the length of contact $d$ (i.e.\ the characteristic size of the debris particle to be detached), but not on the shape of the debris particle that is detached. The expression of the elastic energy is then

\begin{align}
  E_\mathrm{el} = \frac{B d^2 \tau_\mathrm{j}^2 \mathcal{M} }{ \uppi E' } \textrm{ ,}
  \label{eq:Eel_son}
\end{align}
where $E'=E$ for plane stress and $E'=E/(1-\nu^2)$ for plane strain ($E$ and $\nu$ being the elastic modulus and the Poisson ratio of the material), $B$ is the thickness of the semi-infinite body along the $z$ direction, and $\mathcal{M}$ is an infinite integral term in 2D,

\begin{align}
  \mathcal{M} = \int_0^\infty \frac{ \mathrm{d}y}{y} \textrm{ .}
  \label{eq:M}
\end{align}
The shape of the detached particle still defines the crack path, and in Ref.~\onlinecite{pham2019adhesive} it assumed to be semi-circular. This leads to the adhesive energy

\begin{align}
  E_\mathrm{ad} = \uppi \gamma B d \textrm{ ,}
  \label{eq:Eal_son}
\end{align}
where $\gamma$ is the surface energy of the material, and two semi-circular surfaces of circumference $\uppi d / 2$ are created. The critical length scale is found again by imposing the condition $E_\mathrm{el} = E_\mathrm{ad}$, and it is (for plane stress, as in our simulations)

\begin{align}
  d^*_\mathrm{flat} = \frac {\uppi^2 \gamma E }{ \tau_\mathrm{j}^2 \mathcal{M} } \textrm{ .}
  \label{eq:dstar_son}
\end{align}
As $\Delta w = 4\gamma$ and, for the model potentials adopted in Ref.~\onlinecite{aghababaei2016critical} and in this work, $E/G = 8/3$, we can express $d^*_\mathrm{flat}$ in terms of $d^*$:

\begin{align}
  d^*_\mathrm{flat} = \frac{2}{3} \frac{\uppi^2}{\mathcal{M}} \frac{w}{\tau_\mathrm{j}^2/2G} = \frac{2}{3} \frac{\uppi^2}{\mathcal{M}} {} \frac{d^*}{\Lambda} \textrm{ .}
\end{align}
The integral $\mathcal{M}$ is finite in our case, as it is bounded by the plastic zone ($\approx 1$~$r_0$), where the integral is capped ($\int_0^{r_0} \mathrm{d}y/r_0=1$) and the height of the body along the $y$ direction ($\approx 150$~$r_0$). It is then $\mathcal{M} \approx 6$, and

\begin{align}
  d^*_\mathrm{flat} = \frac{\uppi^2}{9} {} \frac{d^*}{\Lambda} \textrm{ .}
  \label{eq:dstar_son_ramin}
\end{align}
which is true if $\Lambda = \uppi^2/9 \approx 1.10$. In the flat contact case, $\Lambda$ is then smaller than the empirical value of $\Lambda_\mathrm{max}=1.50$ found in Ref.~\onlinecite{aghababaei2016critical} for the asperity case.

In the simulations of this current work, the first contact takes place between two rough self-affine surfaces, for which the concept of asperity is ill-defined. Furthermore, when the root mean square of heights $\sigma$ of the surfaces is small, the collision between the two bodies is close to the flat contact situation. This is shown in Supplementary Figure~\ref{fig:lambda}, where it is also observed that the areas that contribute to the detached particle are not semicircular (red atoms in Supplementary Figure~\ref{fig:lambda}(d)). We thus derive $d^*_\mathrm{flat}$ for the general case of a semi-elliptical detached particle. Note that, as previously explained, the elastic energy $E_\mathrm{el}$ contained in the body is independent of the shape of the detached particle, and we need to generalize only the expression for $E_\mathrm{ad}$.

As no exact formula for the perimeter of an ellipse exists, we rely on the second Ramanujan approximation, which provides an accurate estimation also in the limiting case $a/b \to 0$ (two overlapping segments), $2a$ and $2b$ being the axes of the ellipse. The formula gives the perimeter $P(a,b)$ as

\begin{align}
  P(a,b) = \uppi \left( a+b \right) \left( 1 + \frac{3h(a,b)}{10 + \sqrt{4-3h(a,b)} } \right) \textrm{ ,}
  \label{eq:P}
\end{align}
where

\begin{align}
  h(a,b) = \frac{ \left( a - b \right)^2 }{ \left( a + b \right)^2 } \textrm{ .}
  \label{eq:h}
\end{align}

We substitute for $a=d/2$ and $b=\kappa d/2$, where we assume that the major axis is the contact junction $d$ and that the proportionality of the minor axis to the major axis is given by the scalar $0 < \kappa \le 1$. Expressions~\ref{eq:P} and~\ref{eq:h} become

\begin{align}
  P(\kappa, d) = \uppi \frac{d}{2} \left( 1+\kappa \right) \left( 1 + \frac{3h(\kappa,d)}{10 + \sqrt{4-3h(\kappa,d)} } \right) = d {} p\left(\kappa \right) \\
  h(\kappa,d) = \frac{ \left( \frac{1}{2} - \frac{\kappa}{2} \right)^2 }{ \left( \frac{1}{2} + \frac{\kappa}{2} \right)^2 } \textrm{ ,}
\end{align}
where

\begin{align}
  p(\kappa) = \frac{\uppi}{2} \left( 1+\kappa \right) \left( 1 + \frac{3h(\kappa,d)}{10 + \sqrt{4-3h(\kappa,d)} } \right) \textrm{ .}
\end{align}

The adhesive energy (Eq.~\ref{eq:Eal_son}) is now expressed as

\begin{align}
  E_\mathrm{ad} = \gamma B d p(\kappa)  \textrm{ ,}
  \label{eq:Eal_k}
\end{align}
and the critical length scale $d^*_\mathrm{flat}$ (Eq.~\ref{eq:dstar_son}) becomes

\begin{align}
  d^*_\mathrm{flat} \left( \kappa \right) = \frac {\uppi p(\kappa) \gamma E }{ \tau_\mathrm{j}^2 \mathcal{M} } = \frac { \uppi p(\kappa) }{ 9 } \frac{d^*}{\Lambda} \textrm{ .}
  \label{eq:lambda_flat}
\end{align}

The two limit cases are $\kappa = 1$ (circular particle), for which $p(\kappa)=\uppi$ and Equations~\ref{eq:dstar_son} and~\ref{eq:dstar_son_ramin} are recovered, and $\kappa \to 0$ (two overlapping segments), for which $p(\kappa) = 2$, and is

\begin{align}
  d^*_\mathrm{flat} \left(\kappa \to 0 \right) = \frac {2 \uppi \gamma E }{ \tau_\mathrm{j}^2 \mathcal{M} } = \frac{2\uppi}{9} {} \frac{d^*}{\Lambda} \textrm{ ,}
\end{align}
and $\Lambda = 0.70 = \Lambda_\mathrm{min}$.

Supplementary Figure~\ref{fig:lambda}(e) shows the variation of $\Lambda$ as a function of the factor $\kappa$, and it grows sub-linearly for small values of $\kappa$, i.e.\ a high eccentricity of the ellipse gives values of $\Lambda$ close to $\Lambda_\mathrm{min}$.

In the case of the simulations of Supplementary Figure~\ref{fig:lambda}(a-d), we can derive the minor and major axes $2a$ and $2b$ from the position at the first contact of the atoms that will later form the debris particle (Supplementary Figure~\ref{fig:lambda}(d)). The atoms describe two distinct portions, one on each surfaces, whose maximum and minimum lengths are estimated. The average of the two minimum (maximum) lengths gives us an approximation of the axis $2a$ ($2b$), and, thus, $\kappa=0.23$. We can then compute $\Lambda$ from Eq.~\ref{eq:lambda_flat}, which gives $\Lambda=0.74$.

\clearpage
\subsection{Supplementary figures}

\setcounter{figure}{0}
\renewcommand{\thefigure}{S.\hspace{0.05em}\arabic{figure}}

\begin{figure*}[h]
  \centering
  \includegraphics[width=.8\textwidth]{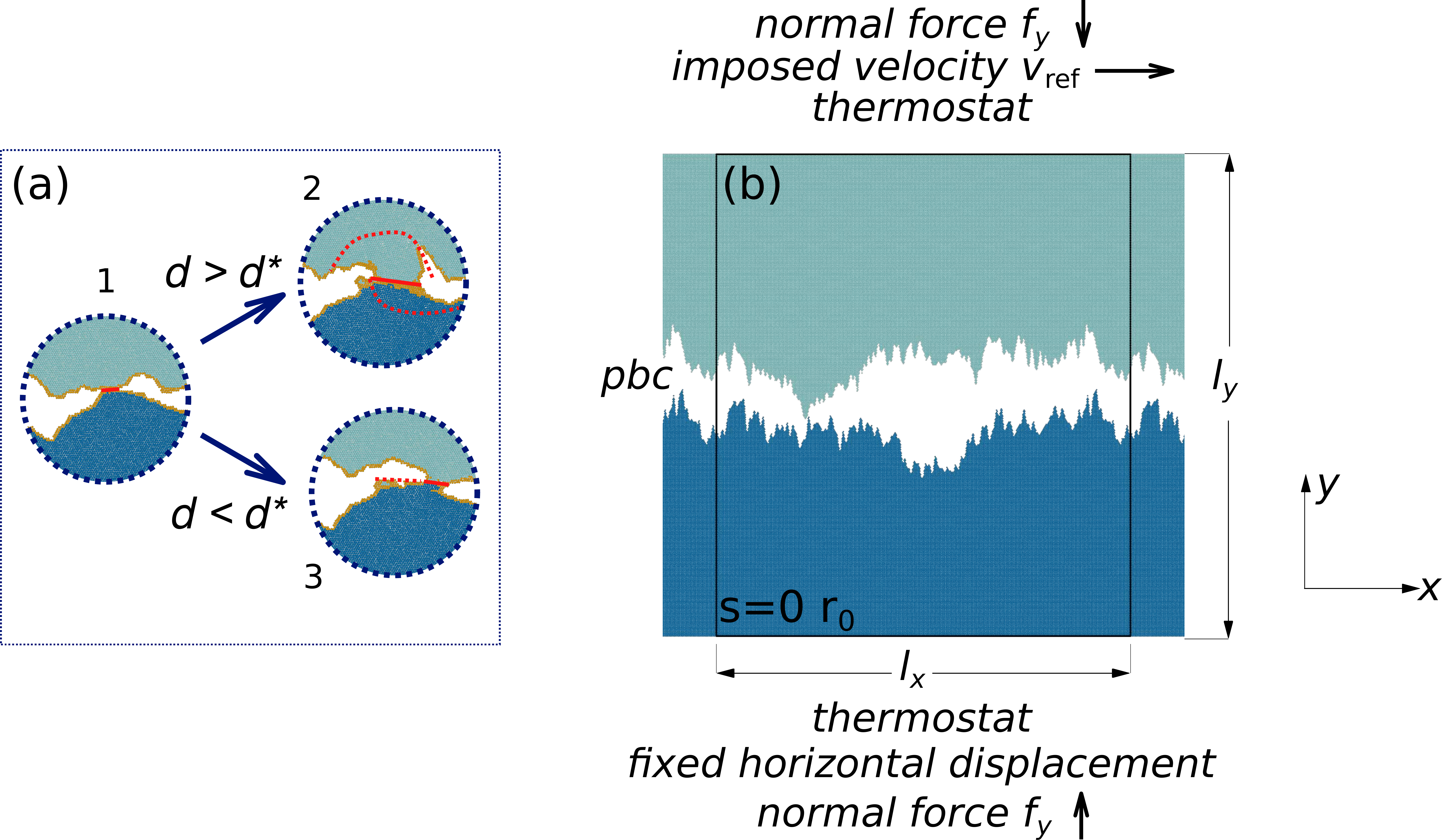}
  \caption{Ductile-to-brittle transition and simulation setup. (a) When two asperities collide, (1), a junction of size $d$ is formed - if it is larger than the critical, material-dependent value $d^*$, cracks appear at both surfaces and a wear debris particle is formed (2), otherwise asperities smooth each other (3). Solid red lines represent the junction of size $d$ and dotted red lines represent the crack path in (2) and the sliding distance in (3). (b) Setup for both sets S and L. The bottom body has zero horizontal velocity, its first layer of atoms being fixed horizontally. The top body slides against the bottom one with velocity $v_\mathrm{ref}$, which is imposed on the top layer of atoms. The normal force $f_y$ pushes the two bodies against one another to ensure contact. Periodic boundary conditions are enforced along $x$, and the simulation box can expand and shrink along $y$. A thermostat in each body is applied on the layers next to the fixed boundaries. In all panels colors distinguish atoms originally belonging to the top (light blue) and bottom (dark blue) bodies. Atoms that at some previous instant were detected as surface atoms and re-assigned to the interfacial adhesion potential are depicted in yellow. In panel (b), black lines represent simulation box boundaries and $s$ is the sliding distance (which is zero).}
  \label{fig:setup}
\end{figure*}

\begin{figure*}[h]
  \centering
  \includegraphics[width=0.35\textwidth]{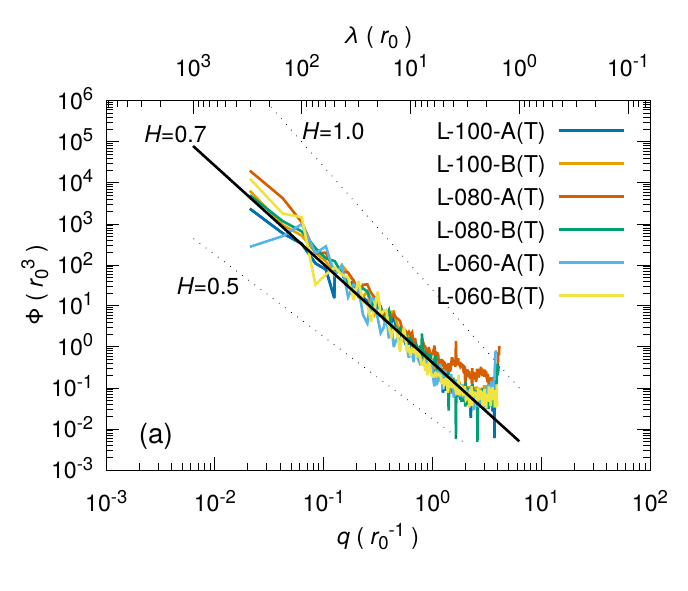}
  \includegraphics[width=0.35\textwidth]{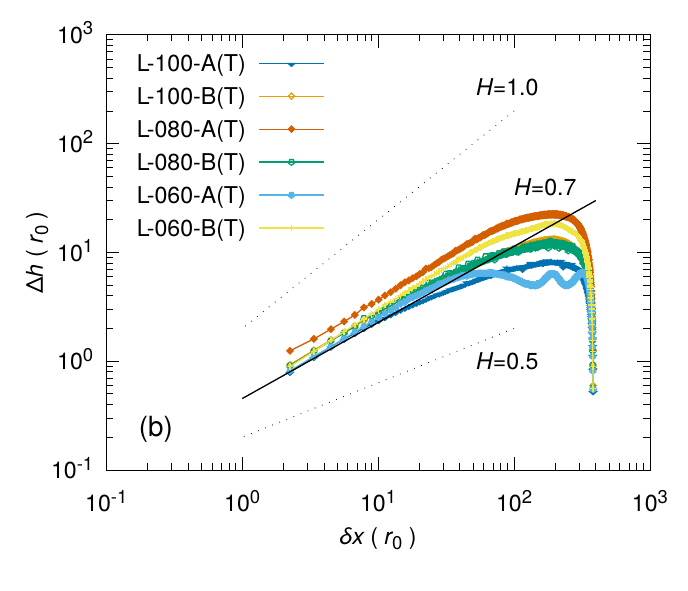}
  \includegraphics[width=0.35\textwidth]{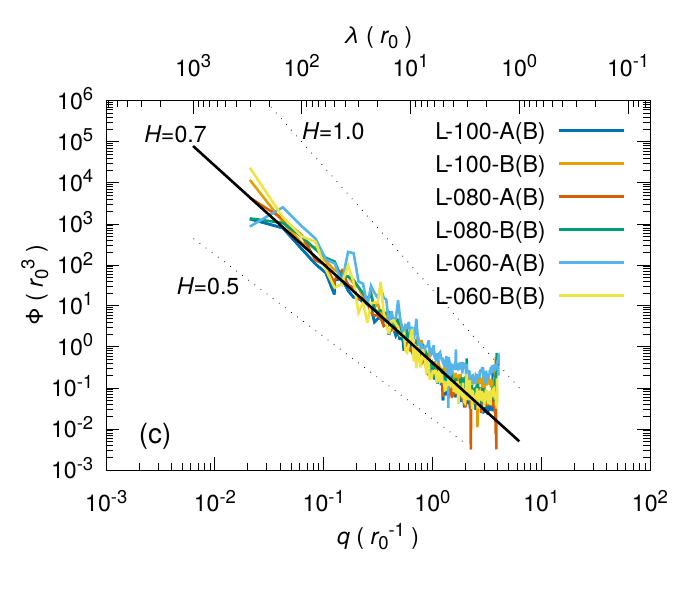}
  \includegraphics[width=0.35\textwidth]{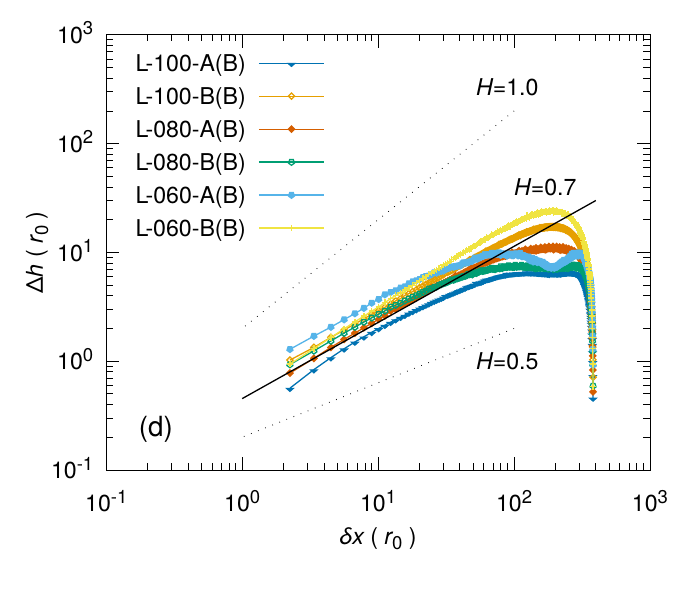}
  \includegraphics[width=0.35\textwidth]{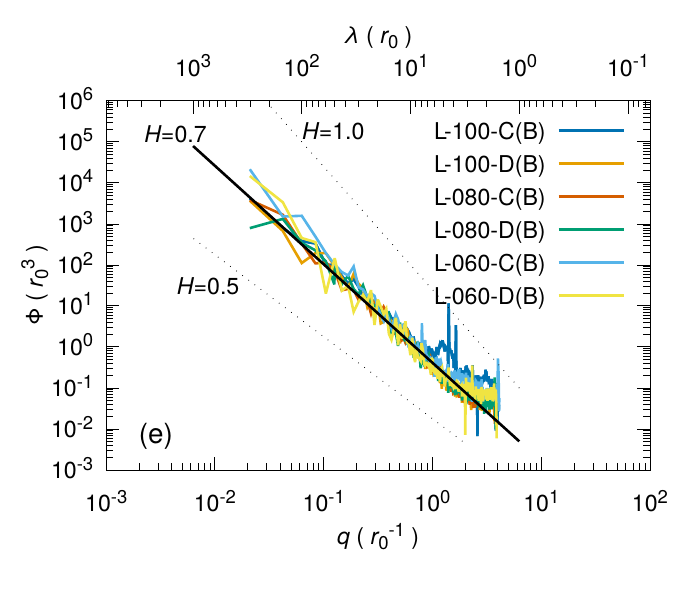}
  \includegraphics[width=0.35\textwidth]{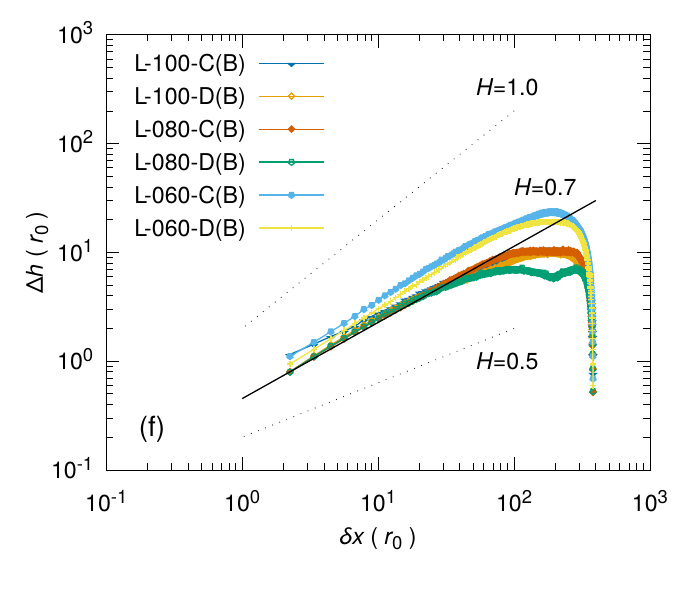}
  \includegraphics[width=0.35\textwidth]{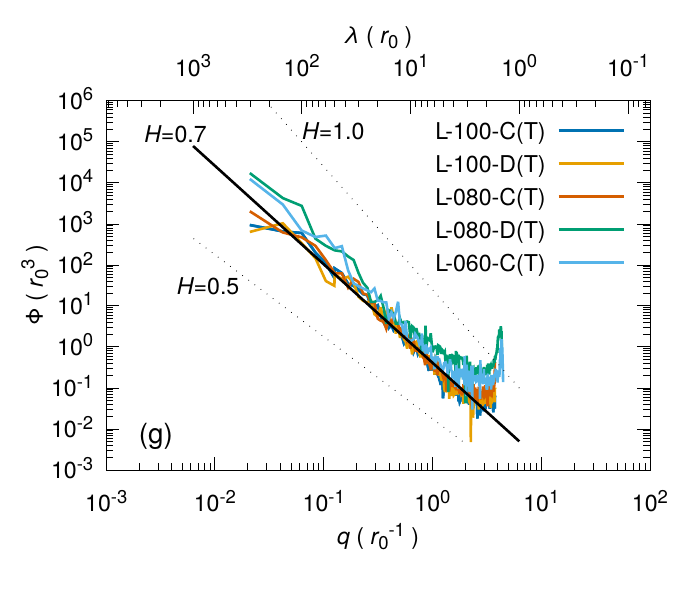}
  \includegraphics[width=0.35\textwidth]{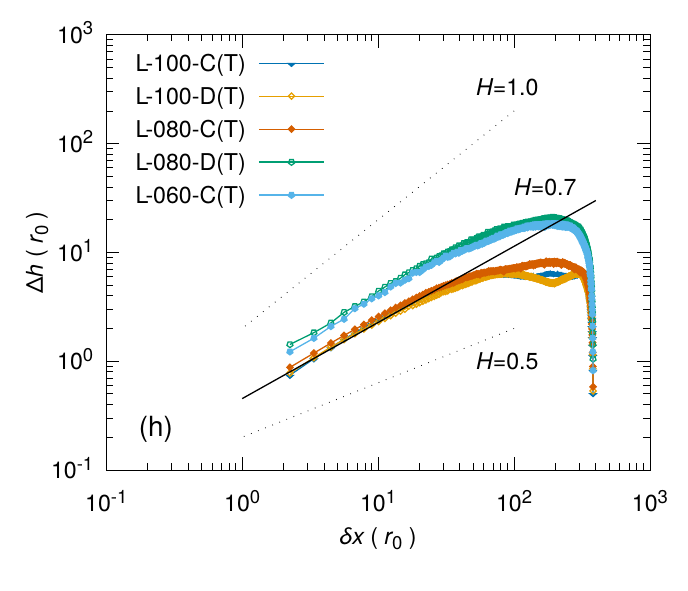}
  \caption{Steady-state surface morphology analysis. Left panels: PSD per unit length $\Phi$ as a function of the wavevector $q$ and the wavelength $\lambda$, the relation between the two being $q=2\uppi/\lambda$. Right panels: height-height correlation function $\Delta h (\delta x) = \langle \left[ h(x+\delta x) - h(x) \right]^2 \rangle ^{1/2}$. The surfaces are taken from the bottom ('(B)') and top ('(T)') bodies of different simulations with different values of interfacial adhesion $\tilde{\gamma}$ and system sizes (see Supplementary Table~\ref{tab:L} for details). Further surfaces are reported in Figure~\ref{fig:H030b}. In all panels the solid black straight guide-line corresponds to a Hurst exponent $H=0.7$. Dotted black straight guide-lines show the hypothetical slope for distributions of $H=0.5$ and $H=1.0$.}
  \label{fig:H}
\end{figure*}

\begin{figure*}
  \centering
  \includegraphics[width=\textwidth]{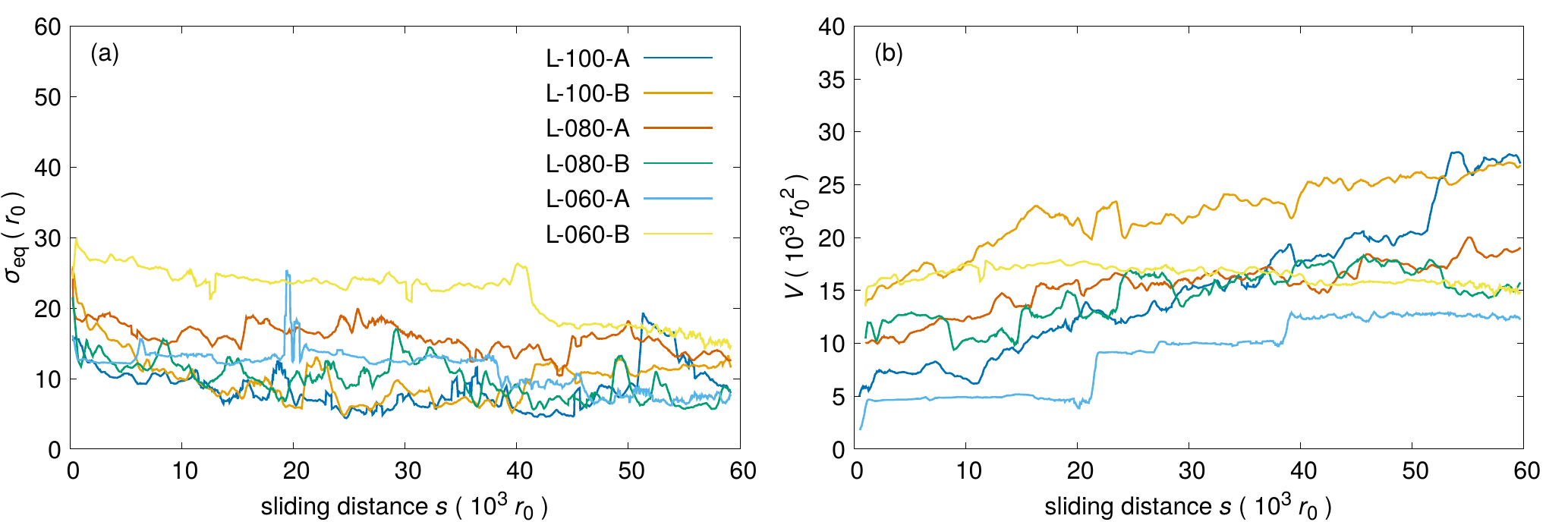}
  \caption{Evolution of the equivalent roughness $\sigma_\mathrm{eq}$ and of the wear volume $V$. (a) Evolution of $\sigma_\mathrm{eq}$ for the simulations in set L (see Supplementary Table~\ref{tab:S10}). For simulations with $\tilde{\gamma}=1.0$ and $\tilde{\gamma}=0.8$, $\sigma_\mathrm{eq}$ decreases until a steady-state is reached, where possible fluctuations due to local events can take place. For simulations with $\tilde{\gamma}=0.6$, the steady-state is not always reached, because of the long sticking times (see also Figure~\ref{fig:H030_travel}). (b) Evolution of the wear volume $V$ of the rolling debris particle, as defined only after its formation. The simulations with $\tilde{\gamma}=1.0$ display steady growth of the particle volume, while simulations with $\tilde{\gamma}=0.6$ are characterized by a negligible wear rate for most of the sliding distance, and the wear volume significantly increases only through fracture-like brittle events (see Figure~\ref{fig:H030_travel}). Simulations with $\tilde{\gamma}=0.8$ show an intermediate behavior.}
  \label{fig:rms_supp}
\end{figure*}

\begin{figure*}[h]
  \centering
  \includegraphics[width=0.49\textwidth]{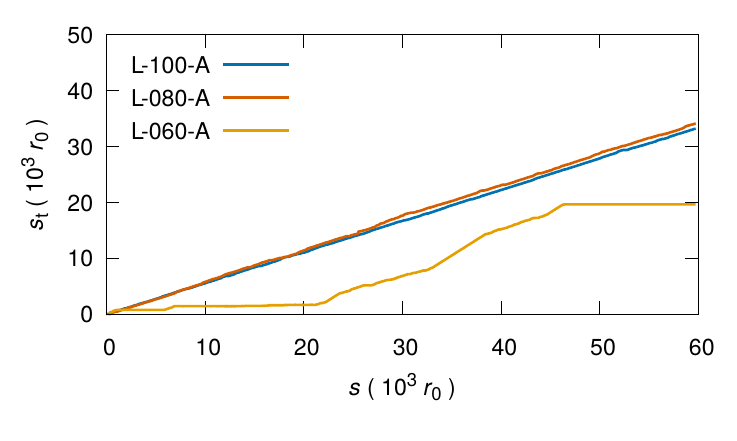}
  \includegraphics[width=0.49\textwidth]{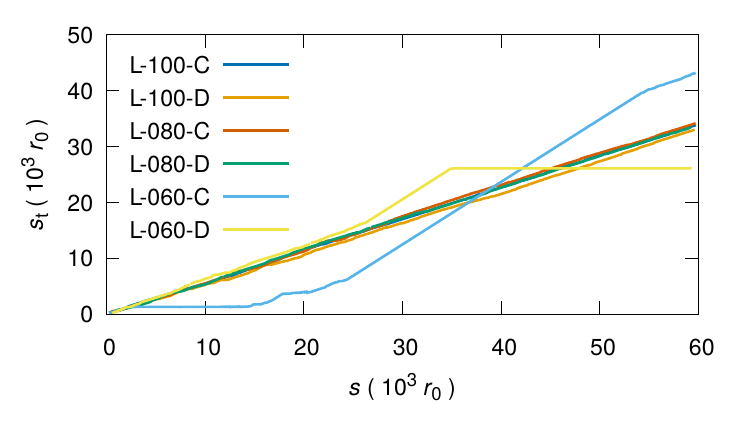}
  \caption{Distance $s_\mathrm{t}$ travelled by the debris particle as a function of the sliding distance. For full and intermediate adhesion simulations (i.e.\ $\tilde{\gamma}=1.0$ and $\tilde{\gamma}=0.8$), the particle rolls most of the time. For low adhesion cases ($\tilde{\gamma}=0.6$), the particle undergoes long times of sticking to one surface (and sliding against the other). These periods are characterized by $\Delta s_\mathrm{t} / \Delta s \approx 0.0$ and $\Delta s_\mathrm{t} / \Delta s \approx 1.0$. See Figure~\ref{fig:H030_travel} for data from further simulations.}
  \label{fig:travel_supp}
\end{figure*}

\begin{figure*}[h]
  \centering
  \includegraphics[width=0.6\textwidth]{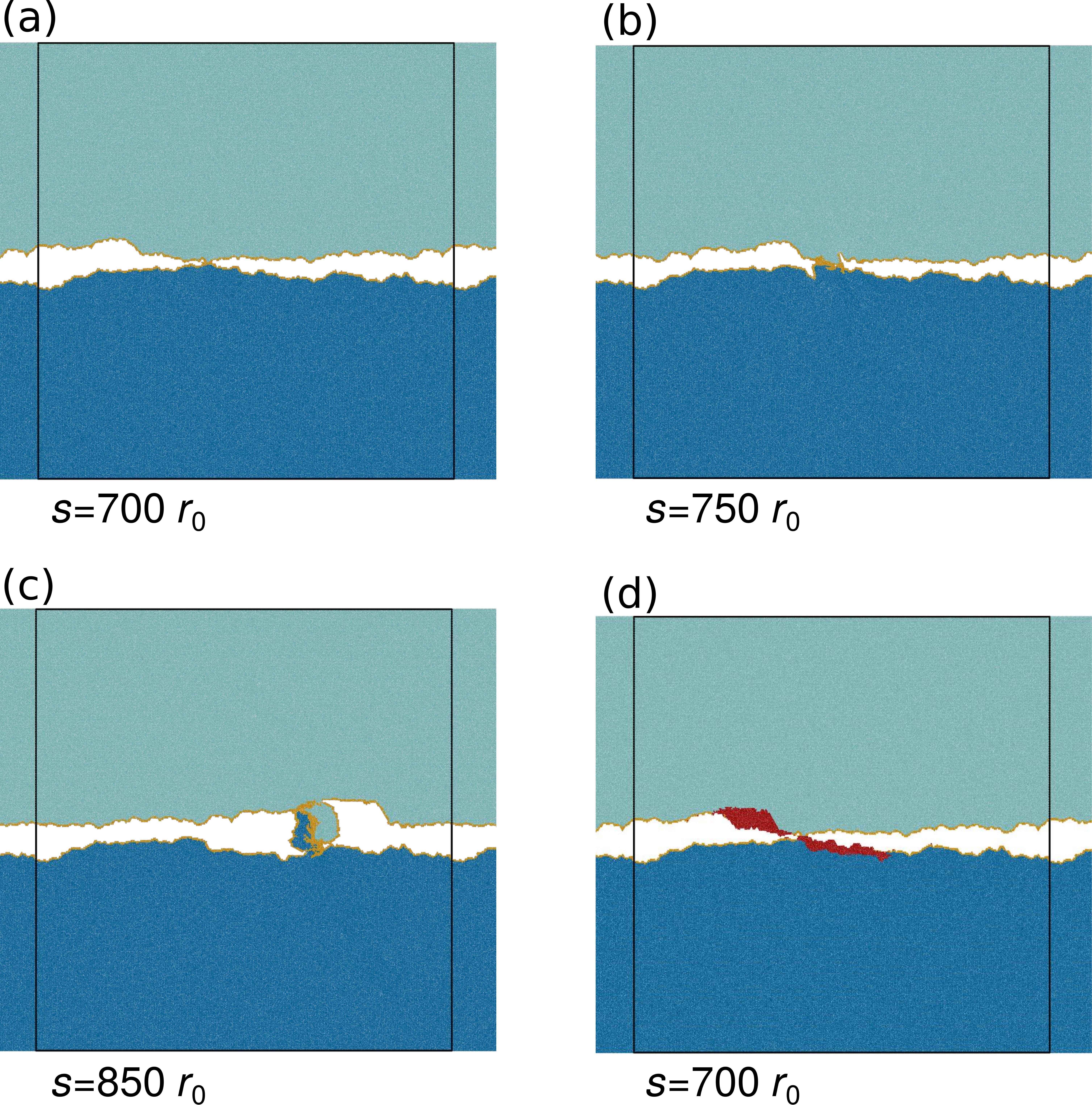}
  \includegraphics[width=0.6\textwidth]{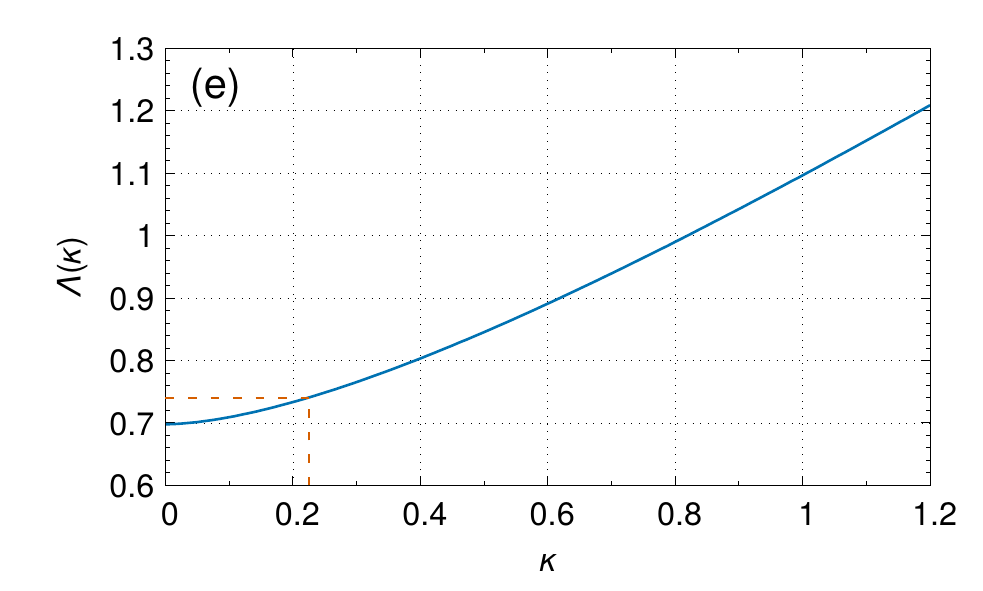}
  \caption{Effect of roughness on debris particle formation and $d^*$. (a-d) The top surface slides and comes into contact with the bottom surface over a few atoms (a). Upon further sliding, the junction grows until cracks on both surfaces appear (b), and a debris particle is finally formed (c). The positions at $s=700$~$r_0$ of the atoms belonging to the debris particle are highlighted in (d). In the case of rough surfaces, the initial geometry affect the stress distribution and the crack path differently then in the well-defined asperities case. This is highlighted by the distribution in the original surface of the atoms that later belong to the debris particle (d). The detached debris particle involves atoms close to the surface, and the process resembles the flat contact case~\cite{pham2019adhesive}. This implies a smaller geometrical factor $\Lambda$ in the expression of $d^*$ than the well-defined asperities case~\cite{aghababaei2016critical} (see Supplementary Methods). (e) Values of the geometrical factor $\Lambda$ as a function of the shape of the detached particle for the simulations of the present work. In the case of the simulation of this Figure, the average ratio between the minor and major sizes of the two red areas (d) is $\kappa = 0.23$, which gives $\Lambda = 0.74$ (dashed orange lines). In panels (a-d) colors distinguish atoms originally belonging to the top (light blue) and bottom (dark blue) surfaces. Atoms that at some previous instant were detected as surface atoms and re-assigned to the interfacial adhesion potential are depicted in yellow. In panel (d), dark red atoms identify atoms that belong to the debris particle shown in panel (c). Black lines represent simulation box boundaries and $s$ is the sliding distance expressed in units of $r_0$. Snapshots in are from simulation S-100-03 (see Supplementary Table~\ref{tab:S10}).}
  \label{fig:lambda}
\end{figure*}

\clearpage
\subsection{Supplementary Tables}

\setcounter{table}{0}
\renewcommand{\thetable}{S.\hspace{0.05em}\Roman{table}}

\begin{table*}[h]
\centering
\begin{tabular}{|P{1.5cm}|P{1.5cm}|P{1.5cm}|P{1.5cm}|P{1.5cm}|P{1.5cm}|P{1.5cm}|}
\hline
 ID & $\tilde{\gamma}$ & $H$ & $\sigma$~$(r_0)$ & seed & scenario & $V_0$~$(10^3 r_0^2)$ \\ \hline
S-100-01 & 1.00 & 0.5 & 5 & 19 & 1 & 6.50 \\ \hline
S-100-02 & 1.00 & 0.5 & 5 & 32 & 1 & 7.34 \\ \hline
S-100-03 & 1.00 & 0.5 & 5 & 42 & 1 & 0.97 \\ \hline
S-100-04 & 1.00 & 0.5 & 5 & 66 & 2 & n/a \\ \hline
S-100-05 & 1.00 & 0.5 & 5 & 91 & 1 & disc. \\ \hline
S-100-06 & 1.00 & 0.7 & 5 & 19 & 1 & 9.32 \\ \hline
S-100-07 & 1.00 & 0.7 & 5 & 32 & 1 & 11.19 \\ \hline
S-100-08 & 1.00 & 0.7 & 5 & 42 & 2 & n/a \\ \hline
S-100-09 & 1.00 & 0.7 & 5 & 66 & 2 & n/a \\ \hline
S-100-10 & 1.00 & 0.7 & 5 & 91 & 2 & n/a \\ \hline
S-100-11 & 1.00 & 1.0 & 5 & 19 & 1 & 8.53 \\ \hline
S-100-12 & 1.00 & 1.0 & 5 & 32 & 1 & 8.55 \\ \hline
S-100-13 & 1.00 & 1.0 & 5 & 42 & 2 & n/a \\ \hline
S-100-14 & 1.00 & 1.0 & 5 & 66 & 1 & 40.34 \\ \hline
S-100-15 & 1.00 & 1.0 & 5 & 91 & 2 & n/a \\ \hline
S-100-16 & 1.00 & 0.5 & 10 & 19 & 2 & n/a \\ \hline
S-100-17 & 1.00 & 0.5 & 10 & 32 & 2 & n/a \\ \hline
S-100-18 & 1.00 & 0.5 & 10 & 42 & 2 & n/a \\ \hline
S-100-19 & 1.00 & 0.5 & 10 & 66 & 1 & disc. \\ \hline
S-100-20 & 1.00 & 0.5 & 10 & 91 & 1 & 11.68 \\ \hline
S-100-21 & 1.00 & 0.7 & 10 & 19 & 1 & disc. \\ \hline
S-100-22 & 1.00 & 0.7 & 10 & 32 & 1 & disc. \\ \hline
S-100-23 & 1.00 & 0.7 & 10 & 42 & 2 & n/a \\ \hline
S-100-24 & 1.00 & 0.7 & 10 & 66 & 1 & disc. \\ \hline
S-100-25 & 1.00 & 0.7 & 10 & 91 & 1 & 10.39 \\ \hline
S-100-26 & 1.00 & 1.0 & 10 & 19 & 2 & n/a \\ \hline
S-100-27 & 1.00 & 1.0 & 10 & 32 & 2 & n/a \\ \hline
S-100-28 & 1.00 & 1.0 & 10 & 42 & 2 & n/a \\ \hline
S-100-29 & 1.00 & 1.0 & 10 & 66 & 1 & disc. \\ \hline
S-100-30 & 1.00 & 1.0 & 10 & 91 & 1 & 11.24 \\ \hline
S-100-31 & 1.00 & 0.5 & 20 & 19 & 1 & 11.13 \\ \hline
S-100-32 & 1.00 & 0.5 & 20 & 32 & 1 & disc. \\ \hline
S-100-33 & 1.00 & 0.5 & 20 & 42 & 2 & n/a \\ \hline
S-100-34 & 1.00 & 0.5 & 20 & 66 & 1 & 11.08 \\ \hline
S-100-35 & 1.00 & 0.5 & 20 & 91 & 1 & 2.20 \\ \hline
S-100-36 & 1.00 & 0.7 & 20 & 19 & 1 & 20.25 \\ \hline
S-100-37 & 1.00 & 0.7 & 20 & 32 & 1 & 18.49 \\ \hline
S-100-38 & 1.00 & 0.7 & 20 & 42 & 1 & 7.82 \\ \hline
S-100-39 & 1.00 & 0.7 & 20 & 66 & 1 & disc. \\ \hline
S-100-40 & 1.00 & 0.7 & 20 & 91 & 1 & disc. \\ \hline
S-100-41 & 1.00 & 1.0 & 20 & 19 & 1 & 15.28 \\ \hline
S-100-42 & 1.00 & 1.0 & 20 & 32 & 1 & 12.50 \\ \hline
S-100-43 & 1.00 & 1.0 & 20 & 42 & 1 & 2.83 \\ \hline
S-100-44 & 1.00 & 1.0 & 20 & 66 & 1 & 9.74 \\ \hline
S-100-45 & 1.00 & 1.0 & 20 & 91 & 1 & disc. \\ \hline
\end{tabular}
\caption{Summary of the simulations in set S: $\tilde{\gamma}=1.0$. $H$, $\sigma$ and 'seed' are respectively the Hurst exponent, root mean square of heights and random seed used to create the initial self-affine surface. Scenario indicates which of the two situations were observed in the simulations: 1 - a debris particle is formed,  or 2 - no debris particle is generated because a junction comparable to the system size is formed and the surfaces weld or a debris particle of the order of the system size would form. $V_0$ is the initial volume of the debris particle. disc.: detected value discarded because of erroneous measures (see main text for details). n/a: not applicable to simulations that displayed Scenario~2. }
\label{tab:S10}
\end{table*}

\begin{table*}[h]
\centering
\begin{tabular}{|P{1.5cm}|P{1.5cm}|P{1.5cm}|P{1.5cm}|P{1.5cm}|P{1.5cm}|P{1.5cm}|}
\hline
 ID & $\tilde{\gamma}$ & $H$ & $\sigma$~$(r_0)$ & seed & scenario & $V_0$~$(10^3 r_0^2)$ \\ \hline
S-080-01 & 0.80 & 0.5 & 5 & 19 & 2 & n/a \\ \hline
S-080-02 & 0.80 & 0.5 & 5 & 32 & 1 & disc. \\ \hline
S-080-03 & 0.80 & 0.5 & 5 & 42 & 2 & n/a \\ \hline
S-080-04 & 0.80 & 0.5 & 5 & 66 & 2 & n/a \\ \hline
S-080-05 & 0.80 & 0.5 & 5 & 91 & 2 & n/a \\ \hline
S-080-06 & 0.80 & 0.7 & 5 & 19 & 2 & n/a \\ \hline
S-080-07 & 0.80 & 0.7 & 5 & 32 & 1 & 6.63 \\ \hline
S-080-08 & 0.80 & 0.7 & 5 & 42 & 2 & n/a \\ \hline
S-080-09 & 0.80 & 0.7 & 5 & 66 & 2 & n/a \\ \hline
S-080-10 & 0.80 & 0.7 & 5 & 91 & 2 & n/a \\ \hline
S-080-11 & 0.80 & 1.0 & 5 & 19 & 1 & disc. \\ \hline
S-080-12 & 0.80 & 1.0 & 5 & 32 & 2 & n/a \\ \hline
S-080-13 & 0.80 & 1.0 & 5 & 42 & 2 & n/a \\ \hline
S-080-14 & 0.80 & 1.0 & 5 & 66 & 2 & n/a \\ \hline
S-080-15 & 0.80 & 1.0 & 5 & 91 & 2 & n/a \\ \hline
S-080-16 & 0.80 & 0.5 & 10 & 19 & 2 & n/a \\ \hline
S-080-17 & 0.80 & 0.5 & 10 & 32 & 2 & n/a \\ \hline
S-080-18 & 0.80 & 0.5 & 10 & 42 & 2 & n/a \\ \hline
S-080-19 & 0.80 & 0.5 & 10 & 66 & 2 & n/a \\ \hline
S-080-20 & 0.80 & 0.5 & 10 & 91 & 1 & disc. \\ \hline
S-080-21 & 0.80 & 0.7 & 10 & 19 & 2 & n/a \\ \hline
S-080-22 & 0.80 & 0.7 & 10 & 32 & 1 & disc. \\ \hline
S-080-23 & 0.80 & 0.7 & 10 & 42 & 2 & n/a \\ \hline
S-080-24 & 0.80 & 0.7 & 10 & 66 & 1 & 2.61 \\ \hline
S-080-25 & 0.80 & 0.7 & 10 & 91 & 1 & 12.09 \\ \hline
S-080-26 & 0.80 & 1.0 & 10 & 19 & 1 & disc. \\ \hline
S-080-27 & 0.80 & 1.0 & 10 & 32 & 2 & n/a \\ \hline
S-080-28 & 0.80 & 1.0 & 10 & 42 & 2 & n/a \\ \hline
S-080-29 & 0.80 & 1.0 & 10 & 66 & 2 & n/a \\ \hline
S-080-30 & 0.80 & 1.0 & 10 & 91 & 1 & 8.28 \\ \hline
S-080-31 & 0.80 & 0.5 & 20 & 19 & 1 & disc. \\ \hline
S-080-32 & 0.80 & 0.5 & 20 & 32 & 1 & 7.08 \\ \hline
S-080-33 & 0.80 & 0.5 & 20 & 42 & 1 & disc. \\ \hline
S-080-34 & 0.80 & 0.5 & 20 & 66 & 1 & 4.97 \\ \hline
S-080-35 & 0.80 & 0.5 & 20 & 91 & 1 & 3.13 \\ \hline
S-080-36 & 0.80 & 0.7 & 20 & 19 & 1 & 13.70 \\ \hline
S-080-37 & 0.80 & 0.7 & 20 & 32 & 1 & 15.90 \\ \hline
S-080-38 & 0.80 & 0.7 & 20 & 42 & 1 & 0.34 \\ \hline
S-080-39 & 0.80 & 0.7 & 20 & 66 & 1 & 11.53 \\ \hline
S-080-40 & 0.80 & 0.7 & 20 & 91 & 1 & disc. \\ \hline
S-080-41 & 0.80 & 1.0 & 20 & 19 & 1 & 13.53 \\ \hline
S-080-42 & 0.80 & 1.0 & 20 & 32 & 1 & 15.65 \\ \hline
S-080-43 & 0.80 & 1.0 & 20 & 42 & 1 & 4.25 \\ \hline
S-080-44 & 0.80 & 1.0 & 20 & 66 & 1 & 12.04 \\ \hline
S-080-45 & 0.80 & 1.0 & 20 & 91 & 1 & 20.05 \\ \hline
\end{tabular}
\caption{Summary of the simulations in set S: $\tilde{\gamma}=0.8$. $H$, $\sigma$ and 'seed' are respectively the Hurst exponent, root mean square of heights and random seed used to create the initial self-affine surface. Scenario indicates which of the two situations were observed in the simulations: 1 - a debris particle is formed,  or 2 - no debris particle is generated because a junction comparable to the system size is formed and the surfaces weld or a debris particle of the order of the system size would form. $V_0$ is the initial volume of the debris particle. disc.: detected value discarded because of erroneous measures (see main text for details). n/a: not applicable to simulations that displayed Scenario~2. }
\label{tab:S08}
\end{table*}

\begin{table*}[h]
\centering
\begin{tabular}{|P{1.5cm}|P{1.5cm}|P{1.5cm}|P{1.5cm}|P{1.5cm}|P{1.5cm}|P{1.5cm}|}
\hline
 ID & $\tilde{\gamma}$ & $H$ & $\sigma$~$(r_0)$ & seed & scenario & $V_0$~$(10^3 r_0^2)$ \\ \hline
S-060-01 & 0.60 & 0.5 & 5 & 19 & 2 & n/a \\ \hline
S-060-02 & 0.60 & 0.5 & 5 & 32 & 2 & n/a \\ \hline
S-060-03 & 0.60 & 0.5 & 5 & 42 & 2 & n/a \\ \hline
S-060-04 & 0.60 & 0.5 & 5 & 66 & 2 & n/a \\ \hline
S-060-05 & 0.60 & 0.5 & 5 & 91 & 2 & n/a \\ \hline
S-060-06 & 0.60 & 0.7 & 5 & 19 & 2 & n/a \\ \hline
S-060-07 & 0.60 & 0.7 & 5 & 32 & 2 & n/a \\ \hline
S-060-08 & 0.60 & 0.7 & 5 & 42 & 2 & n/a \\ \hline
S-060-09 & 0.60 & 0.7 & 5 & 66 & 2 & n/a \\ \hline
S-060-10 & 0.60 & 0.7 & 5 & 91 & 2 & n/a \\ \hline
S-060-11 & 0.60 & 1.0 & 5 & 19 & 2 & n/a \\ \hline
S-060-12 & 0.60 & 1.0 & 5 & 32 & 2 & n/a \\ \hline
S-060-13 & 0.60 & 1.0 & 5 & 42 & 2 & n/a \\ \hline
S-060-14 & 0.60 & 1.0 & 5 & 66 & 2 & n/a \\ \hline
S-060-15 & 0.60 & 1.0 & 5 & 91 & 2 & n/a \\ \hline
S-060-16 & 0.60 & 0.5 & 10 & 19 & 2 & n/a \\ \hline
S-060-17 & 0.60 & 0.5 & 10 & 32 & 2 & n/a \\ \hline
S-060-18 & 0.60 & 0.5 & 10 & 42 & 2 & n/a \\ \hline
S-060-19 & 0.60 & 0.5 & 10 & 66 & 2 & n/a \\ \hline
S-060-20 & 0.60 & 0.5 & 10 & 91 & 1 & 5.47 \\ \hline
S-060-21 & 0.60 & 0.7 & 10 & 19 & 1 & 7.81 \\ \hline
S-060-22 & 0.60 & 0.7 & 10 & 32 & 2 & n/a \\ \hline
S-060-23 & 0.60 & 0.7 & 10 & 42 & 2 & n/a \\ \hline
S-060-24 & 0.60 & 0.7 & 10 & 66 & 2 & n/a \\ \hline
S-060-25 & 0.60 & 0.7 & 10 & 91 & 1 & 7.18 \\ \hline
S-060-26 & 0.60 & 1.0 & 10 & 19 & 1 & disc. \\ \hline
S-060-27 & 0.60 & 1.0 & 10 & 32 & 2 & n/a \\ \hline
S-060-28 & 0.60 & 1.0 & 10 & 42 & 2 & n/a \\ \hline
S-060-29 & 0.60 & 1.0 & 10 & 66 & 2 & n/a \\ \hline
S-060-30 & 0.60 & 1.0 & 10 & 91 & 1 & 4.08 \\ \hline
S-060-31 & 0.60 & 0.5 & 20 & 19 & 1 & 10.45 \\ \hline
S-060-32 & 0.60 & 0.5 & 20 & 32 & 1 & 10.42 \\ \hline
S-060-33 & 0.60 & 0.5 & 20 & 42 & 1 & 2.14 \\ \hline
S-060-34 & 0.60 & 0.5 & 20 & 66 & 1 & disc. \\ \hline
S-060-35 & 0.60 & 0.5 & 20 & 91 & 1 & 0.85 \\ \hline
S-060-36 & 0.60 & 0.7 & 20 & 19 & 1 & disc. \\ \hline
S-060-37 & 0.60 & 0.7 & 20 & 32 & 1 & 13.05 \\ \hline
S-060-38 & 0.60 & 0.7 & 20 & 42 & 1 & 2.04 \\ \hline
S-060-39 & 0.60 & 0.7 & 20 & 66 & 1 & 3.01 \\ \hline
S-060-40 & 0.60 & 0.7 & 20 & 91 & 1 & disc. \\ \hline
S-060-41 & 0.60 & 1.0 & 20 & 19 & 1 & disc. \\ \hline
S-060-42 & 0.60 & 1.0 & 20 & 32 & 1 & 13.02 \\ \hline
S-060-43 & 0.60 & 1.0 & 20 & 42 & 1 & disc. \\ \hline
S-060-44 & 0.60 & 1.0 & 20 & 66 & 1 & disc. \\ \hline
S-060-45 & 0.60 & 1.0 & 20 & 91 & 1 & 17.67 \\ \hline
\end{tabular}
\caption{Summary of the simulations in set S: $\tilde{\gamma}=0.6$. $H$, $\sigma$ and 'seed' are respectively the Hurst exponent, root mean square of heights and random seed used to create the initial self-affine surface. Scenario indicates which of the two situations were observed in the simulations: 1 - a debris particle is formed,  or 2 - no debris particle is generated because a junction comparable to the system size is formed and the surfaces weld or a debris particle of the order of the system size would form. $V_0$ is the initial volume of the debris particle. disc.: detected value discarded because of erroneous measures (see main text for details). n/a: not applicable to simulations that displayed Scenario~2. }
\label{tab:S06}
\end{table*}

\begin{table*}[h]
\centering
\begin{tabular}{|P{1.5cm}|P{2.5cm}|P{2.5cm}|P{2.5cm}|P{2.5cm}|P{2.cm}|}
\hline
$\tilde{\gamma}$ & \makecell{Scenario~1 \\ $\sigma=5$~$r_0$} & \makecell{Scenario~1 \\ $\sigma=10$~$r_0$} & \makecell{Scenario~1 \\ $\sigma=20$~$r_0$} & \makecell{Scenario~1 \\ Total} & $\overline{V}_0$~$(10^3 r_0^2)$ \\ \hline
$1.0$ & \makecell{ $60.0$~$\%$ \\ $(53.3$~$\%)$} & \makecell{$53.3$~$\%$ \\ $(20.0$~$\%)$} & \makecell{$93.3$~$\%$ \\ $(66.6$~$\%)$} & \makecell{$68.9$~$\%$ \\ $(46.7$~$\%)$} & 13.1 \\ \hline
$0.8$ & \makecell{$20.0$~$\%$ \\ $(6.7$~$\%)$} & \makecell{$40.0$~$\%$ \\ $(20.0$~$\%)$} & \makecell{$100$~$\%$ \\ $(80.0$~$\%)$} & \makecell{$53.3$~$\%$ \\ $(35.6$~$\%)$} & 11.0 \\ \hline
$0.6$ & \makecell{$0.0$~$\%$ \\ $(0.0$~$\%)$} & \makecell{$33.3$~$\%$ \\ $(26.7$~$\%)$} & \makecell{$100$~$\%$ \\ $(60.0$~$\%)$} & \makecell{$44.4$~$\%$ \\ $(28.9$~$\%)$} & 8.6 \\ \hline
all & \makecell{$26.7$~$\%$ \\ $(20.0$~$\%)$} & \makecell{$42.2$~$\%$ \\ $(22.2$~$\%)$} & \makecell{$97.8$~$\%$ \\ $(68.9$~$\%)$} & \makecell{$55.6$~$\%$ \\ $(37.0$~$\%)$} & 10.9 \\ \hline
\end{tabular}
\caption{Summary of the simulations in set S that display Scenario~1. In each cell, the first percentage shows the relative amount of simulations that exhibit Scenario~1, the second percentage, in brackets, indicates the relative amount of simulations that exhibit Scenario~1 and for which the initial volume $V_0$ was not discarded. E.g.\ for $\tilde{\gamma}=1.0$, 9 simulations out of the 15 with $\sigma=5$~$r_0$ displayed Scenario~1 (i.e.\  $60.0$~$\%$), and for 8 out of 15 the initial volume $V_0$ was correctly estimated and taken into account in the analysis (i.e.\  $53.3$~$\%$). 'all' refers to estimation across all the values of $\tilde{\gamma}$. 'Scenario~1 Total' refers to estimation across all the values of $\sigma$. $\overline{V}_0$ is the average initial volume. It emerges that the likelihood of displaying Scenario~1 positively correlates with both $\tilde{\gamma}$ and $\sigma$.}
\label{tab:V0}
\end{table*}

\begin{table*}[h]
\centering
\begin{tabular}{|P{1.5cm}|P{1.5cm}|P{1.5cm}|P{1.5cm}|P{1.5cm}|P{1.5cm}|P{1.5cm}|}
\hline
 ID & $\tilde{\gamma}$ & $H$ & seed & $l_x$ \\ \hline
L-100-A & 1.0 & 0.3 & 29 & 339.314 \\ \hline 
L-100-B & 1.0 & 0.3 & 42 & 339.314 \\ \hline 
L-100-C & 1.0 & 1.0 & 29 & 339.314 \\ \hline 
L-100-D & 1.0 & 1.0 & 42 & 339.314 \\ \hline 
L-080-A & 0.8 & 0.3 & 29 & 339.314 \\ \hline 
L-080-B & 0.8 & 0.3 & 42 & 339.314 \\ \hline 
L-080-C & 0.8 & 1.0 & 29 & 339.314 \\ \hline 
L-080-D & 0.8 & 1.0 & 42 & 339.314 \\ \hline 
L-060-A & 0.6 & 0.3 & 29 & 339.314 \\ \hline 
L-060-B & 0.6 & 0.3 & 42 & 339.314 \\ \hline 
L-060-C & 0.6 & 1.0 & 29 & 339.314 \\ \hline 
L-060-D & 0.6 & 1.0 & 42 & 339.314 \\ \hline 
L-100-X & 1.0 & 0.3 & 42 & 678.627 \\ \hline 
L-080-X & 0.8 & 0.3 & 42 & 678.627 \\ \hline 
\end{tabular}
\caption{Summary of the simulations in set L. 'L' indicates that the set contains long timescale simulations. The three digits that follow represent the value of the interfacial adhesion $\tilde{\gamma}$. $H$ and 'seed' are respectively the Hurst exponent and random seed used to create the initial self-affine surface. In the ID, 'A' and 'B' distinguish two different random realizations of the initial surface for $H=0.3$. Similarly, 'C' and 'D' distinguish two different random realizations of the initial surface for $H=1.0$. 'X' indicates simulations with the largest box size $l_x$.}
\label{tab:L}
\end{table*}

\end{document}